\documentclass[aps,prx,superscriptaddress,amsfonts,amsmath,amssymb,showpacs,floatfix,reprint,longbibliography]{revtex4-1}

\usepackage{url}
\usepackage{bm}
\usepackage{graphicx}
\usepackage{amsmath}
\usepackage{amstext}
\usepackage{amssymb}
\usepackage{amsfonts}
\usepackage{amsbsy}
\usepackage{verbatim}
\usepackage{esvect}
\usepackage{color}
\usepackage[colorlinks=true, urlcolor=blue, linkcolor=blue, citecolor=blue, pdftex]{hyperref}
\usepackage{multirow}
\usepackage{floatrow}
\usepackage{float}
\usepackage{gensymb}
\usepackage{textcomp}
\usepackage{enumitem}
\usepackage[utf8]{inputenc}
\usepackage[version=3]{mhchem}
\usepackage{afterpage}
\usepackage{pbox}
\usepackage{makecell}
\usepackage{array,booktabs}
\usepackage{dsfont}
\usepackage{siunitx}
\usepackage{esvect}
\usepackage{bbold}
\usepackage[T1]{fontenc}
\usepackage{lipsum}

\let\a=\alpha  \let\g=\gamma

   \let\c=\chi

\def\beq{\begin{equation}}
\def\eeq{\end{equation}}
\def\bea{\begin{eqnarray}}
\def\eea{\end{eqnarray}}
\def\ba{\begin{array}}
\def\ea{\end{array}}

\begin{document}
\title{Competing orders in a frustrated Heisenberg model on the Fisher lattice}
\author{Atanu Maity}\email{atanu.maity@iopb.res.in}\affiliation{Institute of Physics, Bhubaneswar 751005, Odisha, India}
\affiliation{Homi Bhabha National Institute, Mumbai 400 094, Maharashtra, India}
\author{Yasir Iqbal}\email{yiqbal@physics.iitm.ac.in}\affiliation{Department of Physics, Indian Institute Of Technology Madras, Chennai 600036, Tamil Nadu, India}
\author{Saptarshi Mandal }\email{saptarshi@iopb.res.in}\affiliation{Institute of Physics, Bhubaneswar 751005, Odisha, India}
\affiliation{Homi Bhabha National Institute, Mumbai 400 094, Maharashtra, India}

\begin{abstract}
We investigate the Heisenberg model on a decorated square (Fisher) lattice in the presence of first neighbor $J_{1}$, second neighbor $J_{2}$, and third neighbor $J_{3}$ exchange couplings, with antiferromagnetic $J_{1}$. The classical ground state phase diagram obtained within a Luttinger-Tisza framework is spanned by two antiferromagnetically ordered phases, and an infinitely degenerate antiferromagnetic chain phase. Employing classical Monte Carlo simulations we show that thermal fluctuations fail to lift the degeneracy of the antiferromagnetic chain phase. Interestingly, the spin wave spectrum of the N\'eel state displays three Dirac nodal loops out of which two are symmetry protected while for the antiferromagnetic chain phase we find symmetry protected Dirac lines. Furthermore, we investigate the spin $S=1/2$ limit employing a bond operator formalism which captures the singlet-triplet dynamics, and find a rich ground state phase diagram host to variety of valence-bond solid orders in addition to antiferromagnetically ordered phases.
\end{abstract}

\date{\today}

\maketitle
\section{Introduction}
In Mott-Hubbard insulators, a reasonable description of the localized electron state at low-temperatures is provided by the Heisenberg spin Hamiltonian~\cite{Heisenberg-1928}. In the presence of frustrated interactions, which could be geometric or parametric in origin, the determination of the ground state and low-energy physics of the Heisenberg model poses itself as a highly nontrivial problem. The principal motivation in the investigation of frustrated spin systems lies in the lure of finding either magnetically ordered ground states with intricate spin textures or highly correlated nonmagnetic phases such as spin liquids~\cite{Pomeranchuk-1941,Balents-2010,Savary-2016,Zhou-2017,Broholm-2020}. To this end, transition metal oxides have attracted much attention as they are found in nature with a rich diversity of geometrically frustrated lattice structures, displaying a wide spectrum of magnetic behaviors~\cite{Rao-1989,Maekawa-1994}. In particular, in one such family of manganese oxide compounds (MnO$_2$) such as K$_{1.5}$(H$_3$O)$_x$Mn$_8$O$_{16}$, Ba$_{1.2}$Mn$_8$O$_{16}$, and $\alpha$\textendash MnO$_2$~\cite{DeGuzman-1994,Ishiwata-2006,Hasegawa-2009}, the Mn ions reside on the vertices of a geometrically frustrated network, namely, the hollandite lattice~\cite{DeGuzman-1994,Suib-1994,Liu-2014}. Experimental studies on these systems have unveiled the presence of a plethora of magnetic phases upon variation of temperature, magnetic field, and doping, which include, an antiferromagnetic state~\cite{Strobel-1984}, a ferromagnet, helimagnetic order~\cite{Sato-1997,Sato-1999}, and spin glass behavior. 

In order to understand the origin of this diversity in magnetic behaviors it is helpful to disentangle the effects of magnetic frustration from those arising due to the presence of impurities. Recently, theoretical studies employing an Ising model on the hollandite lattice~\cite{Crespo-2013a,Crespo-2013b} successfully explained the origin of the antiferromagnetic ground state in the disorder free system~\cite{Yamamoto-1974} as well as the doping-induced transition into a spin-glass state~\cite{Shen-2005,Luo-2009,Luo-2010}. However, the Ising model studied in Ref.~\cite{Crespo-2013a} cannot account for helimagnetic (and in general noncollinear) orders observed in K$_{1.5}$(H$_3$O)$_x$Mn$_8$O$_{16}$ and K$_{0.15}$MnO$_{2}$ at low temperatures~\cite{Sato-1997,Sato-1999}. Experimental investigations on Manganese compounds~\cite{Moussa-1996,Chaboussant-2004,Fabreges-2011} have provided evidence that these systems have small magnetic anisotropies and are thus well described by a Heisenberg model. The zero temperature ($T=0$) classical magnetic phase diagram of the Heisenberg model on the Hollandite lattice allowing for different signs and strengths of nearest-neighbor couplings was studied in Ref.~\cite{Mandal-2014}.

The hollandite lattice can be viewed either as coupled two-dimensional triangular lattices stacked in the $z$-direction or as decorated square lattices (called Fisher lattice) stacked in the $y$-direction~\cite{Mandal-2014}. An understanding of the magnetic Hamiltonian on a lattice which is a two-dimensional projection of the original three-dimensional lattice often provides valuable insights into how magnetic order develops in the original three-dimensional model, and helps flesh out the structure of the (often intricate) spin configurations. In this regard, investigation of the magnetic phases in kagome lattice as an insightful route towards understand the complex magnetism in the pyrochlore lattice is noteworthy~\cite{Lapa-2012,*Fouet-2003,*Iqbal-2019prx}. Herein, we adopt the route of understanding the magnetism of the Hollandite lattice by viewing it as coupled Fisher lattices since the non-trivial mechanism of magnetic order in $\alpha$-MnO$_2$ materials seems to arise due to the coupling in $y$-direction \cite{Mandal-2014,Crespo-2013b,Liu-2014,Amber-2015}. In this work, we carry out a detailed analysis of the magnetic phases present in the $T=0$ classical phase diagram and investigate fluctuation effects beyond the classical limit via a spin-wave analysis and a bond-operator formalism for spin $S=1/2$.

We consider a minimal model on the Fisher lattice [see Fig.~\ref{fig:lattice}] such that $J_1$ couples the vertices of neighboring squares, $J_2$ defines the nearest-neighbour coupling within the squares, and $J_{3}$ is the second nearest-neighbor (diagonal) coupling within each square. The inclusion of a $J_{3}$ coupling has been motivated from recent studies~\cite{Amber-2015,Liu-2014} which suggest that it might be necessary to describe the magnetism in Hollandite systems. At the classical level, a Luttinger-Tisza analysis~\cite{Luttinger-1946,*Luttinger-1951,*Kaplan-2007,*Ghosh-2019} of the $(J_{1}, J_{2},J_{3})$ parameter space reveals the presence of different kinds of antiferromagnetically (AF) ordered states, an infinitely degenerate uncorrelated antiferromagnetic chain phase~\cite{McClarty-2015,Balla-2020}, as well as a unique N\'eel phase which features magnonic Dirac nodal lines  depending on the sign and strength of the couplings. Furthermore, we investigate the role of quantum and thermal fluctuations on these ground states and find via (numerically) exact classical Monte Carlo simulations that thermal fluctuations fail to lift the degeneracy of the uncorrelated antiferromagnetic chain phase. We complement our study by going beyond the spin-wave approximation and compute the relative stability of the semi-classical ground state within a variational ansatz by comparing the energies of competing states and find that each of them is stable as they feature a finite triplon excitation gap over suitable singlet states. 

Our paper is structured as follows. In Sec.~\ref{sec0}, we define the model Hamiltonian and the Luttinger-Tisza framework employed to obtain the classical $T=0$ phase diagram. In Sec.~\ref{grd-st-sec}, we discuss the Luttinger-Tisza ground states and study the effect of thermal fluctuations employing classical Monte Carlo simulations. In Sec.~\ref{spin-wave} and Sec.~\ref{fluctuation}, the impact of quantum fluctuations to harmonic order on the ground states is presented. In Sec.~\ref{vbs}, we analyze our model Hamiltonian for spin $S=1/2$ within the scope of a bond operator formalism and show the existence of three different types of quantum paramagnetic ground states, namely, a plaquette VBS, and two other dimer ordered states. Finally, we summarize and discuss our results in Sec.~\ref{discussion}.

\section{Model and Methods}
\label{sec0}
\begin{figure}[t]
\includegraphics[width=1.0\linewidth]{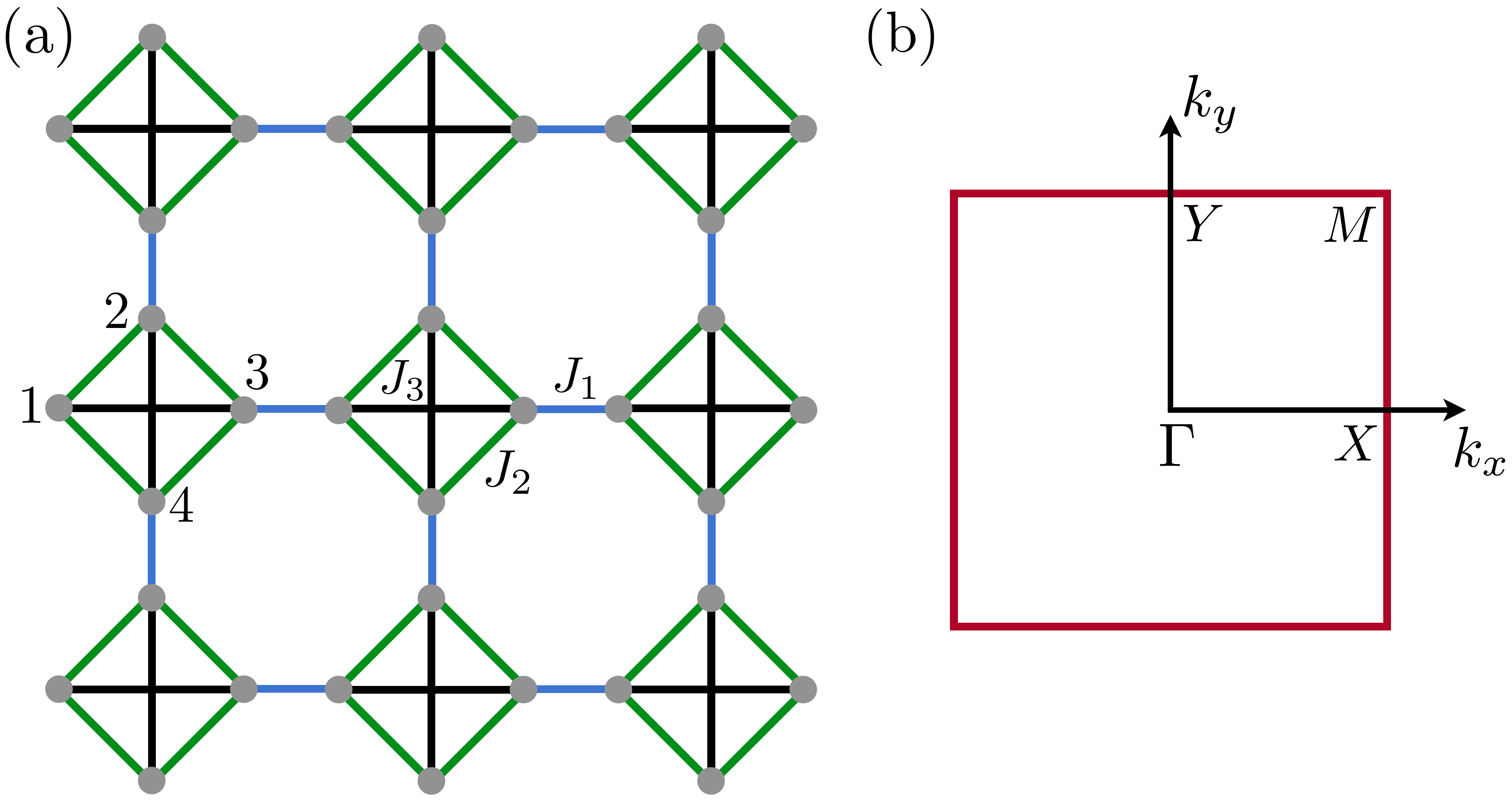}
\caption{(a) The Fisher lattice showing the three exchange couplings $J_1$ (blue) connecting sites on octagons, $J_2$ (green) connecting the sites of a square, and $J_3$ (black) connecting the diagonals of a square, of Eq.~\eqref{heisen}, with the four sites of the unit cell marked by 1, 2, 3, and 4. (b) The first Brillouin zone of the Fisher lattice with the high-symmetry points marked.}
\label{fig:lattice}
\end{figure}
We consider a two-dimensional plane of the Hollandite lattice [see Fig.~\ref{fig:lattice}(a)], called the decorated square (Fisher) lattice, which is characterized by a four-site geometrical unit cell~\footnote{In a Hollandite lattice, the even and odd sublattices lie in different planes, but for the purposes of the present study, this fact is not important.}. The interactions between the spins localized on the vertices of this lattice are governed by a Heisenberg Hamiltonian
\begin{equation}
{\cal \hat{H}}=J_1 \sum_{{\langle i,j \rangle}_{1}} \mathbf{\hat{S}}_{i} \cdot \mathbf{\hat{S}}_{j}
+J_2 \sum_{{\langle i,j \rangle}_{2}} \mathbf{\hat{S}}_{i} \cdot \mathbf{\hat{S}}_{j}+J_3 \sum_{{\langle i,j \rangle}_{3}} \mathbf{\hat{S}}_{i} \cdot \mathbf{\hat{S}}_{j},
\label{heisen}
\end{equation}
where the $J_{1}$, $J_{2}$, and $J_{3}$ superexchange couplings are schematically illustrated in Fig.~\ref{fig:lattice}(a). It is worth noting that in earlier studies~\cite{Crespo-2013a,Mandal-2014} investigating the magnetism of the full three-dimensional Hollandite lattice, the consideration of in-plane interactions was restricted to inter-square $(J_{1})$ and nearest-neighbor intra-square $(J_{2})$ couplings only~\footnote{In Ref.~\cite{Crespo-2013a,Mandal-2014} the $J_{1}$ and $J_{2}$ couplings in the current paper are labelled as $J_{2}$ and $J_{3}$, respectively, while the inter-plane coupling is labelled as $J_{1}$. As we only deal with a given two dimensional plane which is perpendicular to the channel directions, $J_1$ coupling of earlier studies is absent in our present analysis.}, while the inter-plane coupling was found to yield helimagnetic order. Recent experimental studies~\cite{Amber-2015,Liu-2014} on Hollandite compounds have pointed to relatively more intricate ground states compared to those found in Refs.~\cite{Crespo-2013a,Mandal-2014,Sato-1997,Sato-1999}. In particular, in Ref.~\cite{Amber-2015}, the in-plane magnetic ground state was found to possess a magnetic unit cell which is a $4 \times 4$ expansion of the geometrical unit cell. Though the materials in question potentially involve more complex charge orderings which are likely to induce further magnetic coupling between the Mn atoms, it is understood that a simple model which accounts for only the above two mentioned in-plane interactions ($J_{1}$ and $J_{2}$ in Fig.~\ref{fig:lattice}(a)) is not sufficient to explain the formation of a magnetic order with a $4 \times 4$ magnetic unit cell. The above fact motivates us to explore a larger parameter space of exchange couplings, and to this end, we propose the simplest extension by introducing an additional second nearest-neighbor (diagonal) coupling within each square, i.e., $J_{3}$ in Fig.~\ref{fig:lattice}(a). In our study we consider all possible signs and strengths of the $(J_{2},J_{3})$ couplings with an
antiferromagnetic $J_{1}$.

Our analysis of the ground states of the classical version of Eq.~\eqref{heisen} employs the Luttinger-Tisza method. The corresponding classical model is obtained by normalizing the spin operators with respect to their angular momentum $S$ and taking the limit $S\to\infty$~\cite{Millard-1971,Lieb-1973}. Consequently, the spin operators in Eq.~\eqref{heisen} are replaced by ordinary vectors of unit length at each lattice site $i$. For a generic spin interaction, we have the following classical Hamiltonian that needs to be minimized

\begin{equation}\label{eqn:Hamclass}
 {\cal H} = \sum_{i,j, \alpha, \beta} J_{\alpha \beta}(\mathbf{R}_{ij}) \mathbf{S}_{i, \alpha} \cdot \mathbf{S}_{j, \beta},
\end{equation}
where $i,j$ denotes the corresponding Bravais lattice sites separated by lattice translation vectors $\mathbf{R}_{ij}$ and $\alpha,\beta$ indices denote the sublattice sites. The underlying Bravais lattice of the Fisher lattice is the square lattice. The Luttinger-Tisza method~\cite{Luttinger-1946,*Luttinger-1951,*Kaplan-2007} seeks to find a ground state of Eq.~(\ref{eqn:Hamclass}) by enforcing the spin-length constraint at a global level, i.e., $\sum_{i}|\mathbf{S}_{i}^{2}|=NS^{2}$, where $N$ is the total number of lattice sites, a condition termed as the \emph{weak constraint}. This constraint amounts to permitting site-dependent average local moments which take us beyond the classical limit by approximately incorporating some aspects of quantum fluctuations~\cite{Kimchi-2014}.

A solution of this relaxed problem is achieved by decomposing the spin configuration into its Fourier modes $\mathbf{\tilde{S}}_{\alpha}(\mathbf{k})$ on the four sublattices of the Fisher lattice
\begin{equation}
 \mathbf{S}_{i, \alpha} = \frac{1}{\sqrt{N/4}} \sum_{\mathbf{k}} \mathbf{\tilde{S}}_{\alpha}(\mathbf{k}) e^{\imath \mathbf{k} \cdot \mathbf{r}_{i,\alpha}}.
\end{equation}
Inserting this expression into Eq.~(\ref{eqn:Hamclass}) results in
\begin{equation}
 {\cal H} = \sum_{\mathbf{k}}\sum_{\alpha, \beta} \tilde{J}_{\alpha \beta}(\mathbf{k}) \mathbf{\tilde{S}}_{\alpha}(\mathbf{k})\cdot \mathbf{\tilde{S}}_{\beta}(-\mathbf{k}),
\end{equation}
with the interaction matrix given by
\begin{equation}\label{eqn:ltmatrix}
  \tilde{J}_{\alpha \beta}(\mathbf{k}) = \sum_{i,j} J_{\alpha \beta}(\mathbf{R}_{ij}) e^{\imath \mathbf{k} \cdot \mathbf{R}_{ij}}.
\end{equation}

The modes which respect the weak constraint are given by the wave vector $\mathbf{k}$, for which the \emph{lowest} eigenvalue of Eq.~(\ref{eqn:ltmatrix}) has its minimum. The eigenvector corresponding to this eigenvalue gives the relative weight of the modes on the sublattices~\cite{Bertaut-1961}, which means that these modes do not fulfill the strong constraint ($|\mathbf{S}_{i}^{2}|=S^{2}$, i.e., fixed spin-length constraint on every site) if the components of the eigenvector do not have the same magnitude. On the other hand, if this condition is met, the true ground state of the classical model is a coplanar spiral determined by the optimal Luttinger-Tisza wave vector~\cite{Nussinov-2004}.

\section{Classical Ground states}

\subsection{Luttinger-Tisza analysis}\label{grd-st-sec}

The interaction matrix $\tilde{J}_{\alpha \beta}(\mathbf{k})$ for our model takes the form

\begin{widetext}

\[
\begin{pmatrix}
0 & J_{2} e^{\imath(k_{x}-k_{y})a} & J_{3} e^{\imath2k_{x}a}+J_{1}e^{\imath k_{x}b} & J_{2} e^{\imath(k_{x} + k_{y})a}   \\
J_{2} e^{-\imath(k_{x}-k_{y})a} & 0 & J_{2} e^{\imath(k_{x} + k_{y})a} & J_{3} e^{\imath2k_{x}a}+J_{1}e^{\imath k_{x}b} \\
J_{3} e^{-\imath2k_{x}a}+J_{1}e^{-\imath k_{x}b} & J_{2} e^{-\imath(k_{x} + k_{y})a} & 0 & J_{2} e^{\imath(k_{x}-k_{y})a} \\
J_{2} e^{-\imath(k_{x} + k_{y})a}  & J_{3} e^{-\imath2k_{x}a}+J_{1}e^{-\imath k_{x}b} & J_{2} e^{-\imath(k_{x}-k_{y})a} & 0 
\end{pmatrix}
\]

\end{widetext}

In the region of parameter space defined by $J_{3}>|J_{2}|$, we find that the minimal eigenvalue wave vector $(k_x,k_y)_{0}$ is given by
\begin{eqnarray}
(k_{x},k_{y})_{0}&=&(2m\pi/3,k_{y})~~~~~~{\rm or} \notag \\
(k_{x},k_{y})_{0}&=&(k_{x},2n\pi/3)
\end{eqnarray}
where $m,n\in\mathbb{Z}$, hence, realizing long-range ordering in one direction with an absence of relative ordering in the other direction. Along the line $J_{2}=J_{3}$, we find 
\begin{equation}
(k_{x},k_{y})_{0}=(k_x,k_y),
\end{equation}
leading to a degenerate ground state manifold. In the remaining regions of parameter space we find 
\begin{equation}
(k_{x},k_{y})_{0}=((2n+1)\pi/3,(2m+1)\pi/3),
\end{equation} 
which corresponds to long range magnetic order with commensurate ordering wave vectors. The absence of an incommensurate ordering wave vector implies that the degree to which mutual interactions between spins is satisfied is likely to be determined locally, and hence, as a starting point it is helpful to pursue an energy minimization of a local cluster of spins. To this end, we employ a variational approach which proceeds by first constructing spin configurations of a local cluster of spins that minimize its energy and subsequently attempt to construct a global spin configuration which also satisfies the local minimum energy configuration of the cluster of spins. We verify the accuracy of our global spin configurations from classical Monte Carlo simulations. 

\begin{figure}
 \includegraphics[width=1.0\textwidth]{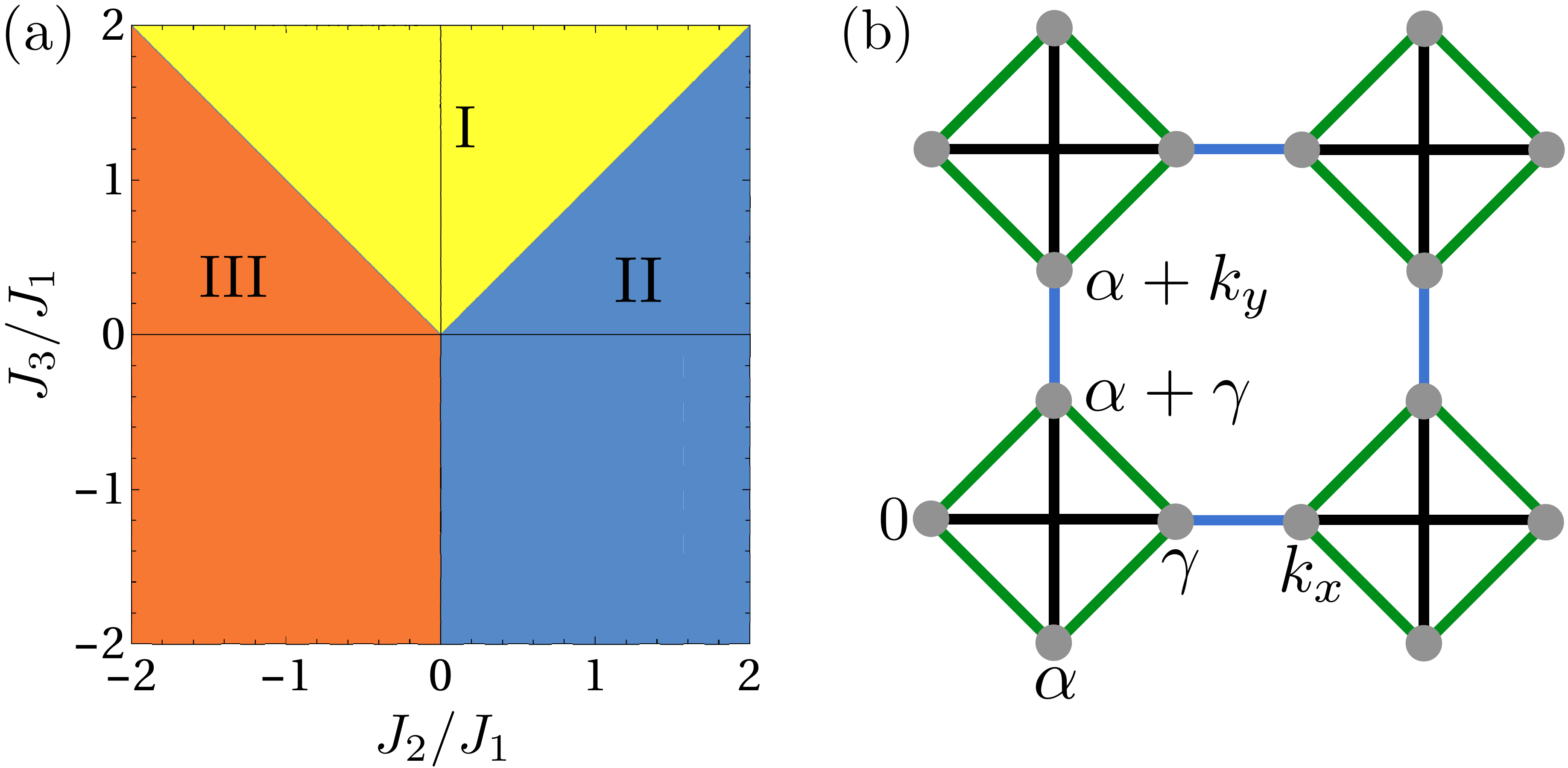}
  \caption{(a) Classical phase diagram of $J_{1}$\textendash$J_{2}$\textendash$J_{3}$ Heisenberg model on the Fisher lattice with the couplings as defined in Fig.~\ref{fig:lattice}(a) and Eq.~\eqref{heisen}, (b) paramterization of a generic spin configuration with $\gamma=k_x+\pi$ and $k_x=k_y=k$.}
\label{fig:classical_phase}
\end{figure}

To start with, we consider a cluster of four spins that constitute a unit cell of the Fisher lattice. The spin configuration is parameterized by observing that the lattice can be described as a collection of horizontal and vertical connections which are coupled via $J_2$ bonds. Each horizontal and vertical string of connections hosts two sublattices each. The relative orientation between the spins within both the sublattices are assigned an angle $\gamma$, while the relative orientation between the spins belonging to the same sublattice but in different chains is assigned an angle $\alpha$ [see Fig.~\ref{fig:classical_phase}(b)]. This choice of ansatz gives an energy density 

\begin{eqnarray}
E/NS^{2} &=& \frac{1}{4}(J_1(\cos(\g -k_x)+\cos(\g-k_y)) + J_2(2\cos \a \nonumber \\ &+& \cos (\a +\g) + \cos (\a -\g) )+2J_3\cos \g).
\label{en_spin}
\end{eqnarray}

The above expression has four free parameters which need to be determined to obtain the ground state spin configuration. We note that since the antiferromagnetic $J_{1}$ bond is not frustrated by any other interaction, one may put forth an ansatz in which the spins connected by the $J_{1}$ bonds are antiparallel, i.e., $k_x-\g=k_y-\g=\pi$, and with Eq.~\eqref{en_spin} simplifying to 

\begin{equation}
E/NS^{2}= \frac{1}{2}(J_2\cos \a (1-\cos k_x)-J_3\cos k_x-J_1).
\label{en_spin_mod}
\end{equation}

Upon minimizing Eq.~\eqref{en_spin_mod} with respect to $k_{x}$ and $\alpha$ we get the following two sets of conditions for a spin configuration to qualify as a ground state
\begin{eqnarray}
\sin k_{x}(J_{2}\cos\alpha+J_{3})&=&0 \label{eq:min1}\\
\sin\alpha(1-\cos k_{x})&=&0.
\label{eq:min2}
\end{eqnarray}
The solutions $(k_{x},\alpha)$ satisfying the conditions [Eqs.~\eqref{eq:min1} and~\eqref{eq:min2}] corresponding to different phases are described below 

\subsubsection{Antiferromagnetic chain phase} 
\begin{figure}
  \includegraphics[width=1.0\textwidth]{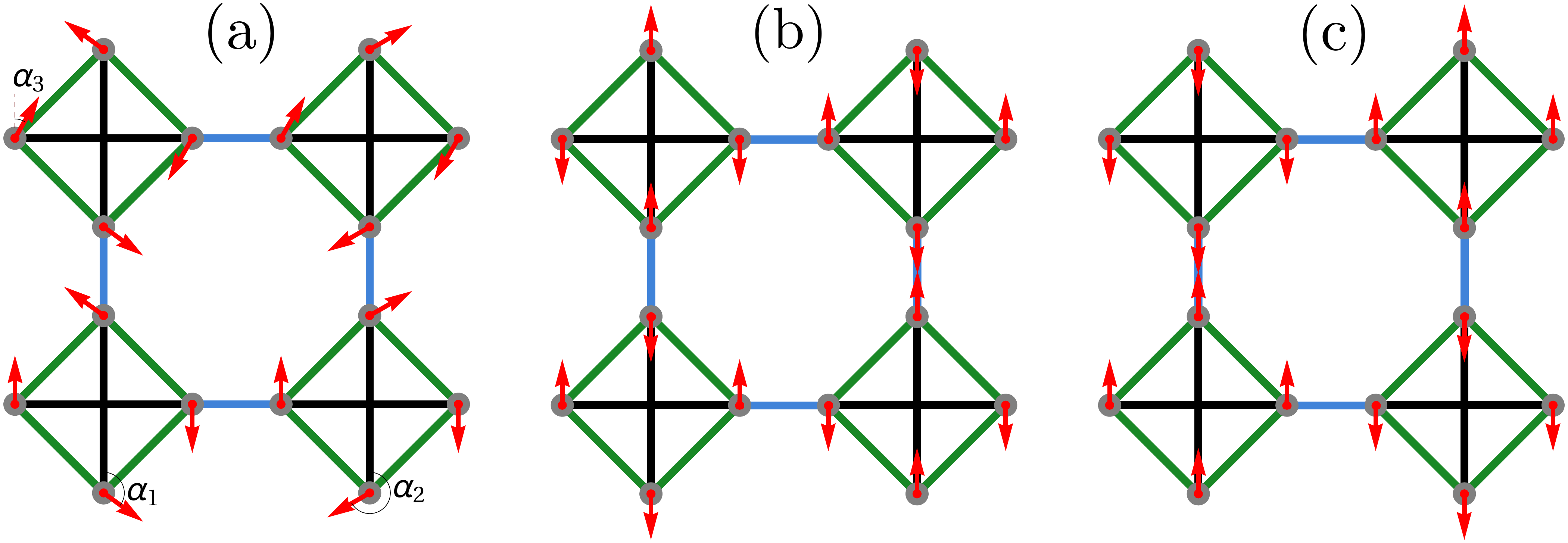}
  \caption{Spin configuration in (a) phase I (uncorrelated antiferromagnetic chain) with $(k,\alpha)=(0,\alpha)$. The individual horizontal or vertical chains have perfect antiferromagnetic order, the relative orientations between them is not fixed. With respect to lower horizontal chain the orientations of the two vertical chains is $\alpha_1$ and $\alpha_2$, respectively. Also, note that the orientation of upper horizontal chain with respect to lower horizontal chain is $\alpha_3$, i.e., the system has an infinite degeneracy, (b) phase II (N\'eel phase) with $(k,\alpha)=(\pi,\pi)$, and (c) phase III (sublattice N\'eel phase) with $(k,\alpha)=(\pi,0)$.}
\label{fig:spinconfigs}
\end{figure}

This phase is characterized by $(k,\alpha)=(0,\alpha)$ where $k=k_{x}=k_{y}$, and is stabilized for $J_3 \geqslant |J_2|$. It is depicted as phase I (yellow region) in the phase diagram of Fig.~\ref{fig:classical_phase}(a). It features perfect antiferromagnetic order along either the horizontal or vertical chains, however, there is a complete absence of spin correlations between any two of these ordered chains [see Fig.~\ref{fig:spinconfigs}(a)]. This implies that within any given four site unit cell, the spins coupled by $J_3$ bonds are antiferromagnetically correlated while there is no correlation between the spins connected by  $J_2$. Hence, the angle $\alpha$ can take any value, implying an infinite degeneracy of the ground state manifold. The ground state energy is then independent of $\alpha$  
\begin{eqnarray}
E/NS^{2}=-\frac{1}{2}(J_3+J_1).
\end{eqnarray}
In the above expression, the independence of the energy on $\alpha$ arises due to a cancellation of the contributions from two bonds connected by $J_2$ within a given square plaquette. The existence of such a degeneracy within each square together with long range antiferromagnetic order along horizontal or vertical chains poses itself as an interesting platform to investigate the order-by-disorder physics driven by thermal and quantum fluctuations. This will be discussed in Sec.~\ref{sec:cmc}, Sec.~\ref{spin-wave} and Sec.~\ref{fluctuation}.

\begin{figure*}
  \includegraphics[width=1.0\textwidth]{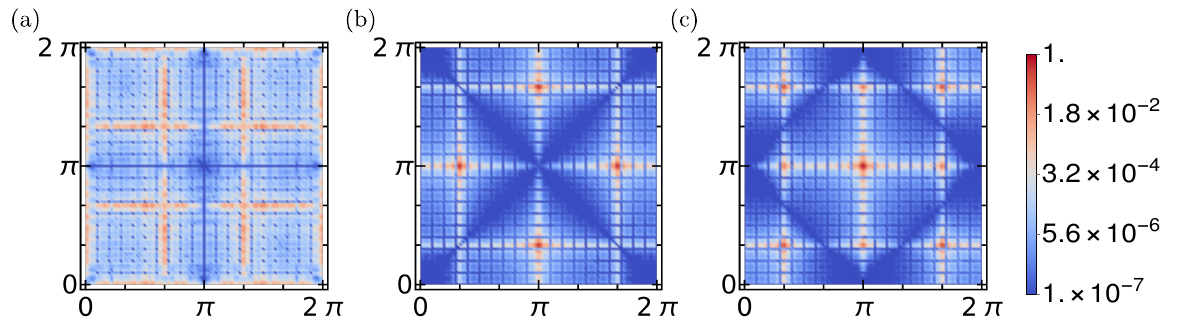}
  \caption{The static (equal-time) spin structure factor obtained from classical Monte Carlo simulations for the three different phases in the classical phase diagram of Fig.~\ref{fig:classical_phase}(a) evaluated at $T/J_{1}S^{2}=0.001$ for a system of 1600 spins, (a) phase I at $(J_{2}/J_{1},J_{3}/J_{1})=(1,4)$, (b) phase II at $(J_{2}/J_{1},J_{3}/J_{1})=(4,1)$,  and (c) phase III $(J_{2}/J_{1},J_{3}/J_{1})=(-4,1)$.}
\label{fig_stf}
\end{figure*}

\subsubsection{N\'eel phase}
For $J_{3}<|J_{2}|$, we enter a region of parameter space where the arbitrariness in the choice of the parameter $\alpha$ in the uncorrelated antiferromagnetic chain phase gets lifted. In particular, for $J_2>0$ and $J_3<J_{2}$, we obtain a N\'eel ordered phase [marked as phase II (blue region) in Fig.~\ref{fig:classical_phase}(a)]. This phase is characterized by $(k,\alpha)=(\pi,\pi)$ which signifies that within any given unit cell (square) there is perfect antiferromagnetic order, and that the spins in neighboring unit cells are aligned antiferromagnetically with respect to each other [see Fig.~\ref{fig:spinconfigs}(b)] . However, unlike the familiar N\'eel phase on the square or honeycomb lattice, not all antiferromagnetic bonds are satisfied when $J_{3}>0$, as the spins connected by the $J_3$ couplings remain frustrated. The ground state energy of this phase is given by, 
\begin{eqnarray}
E/NS^{2}=-\frac{1}{2}(2J_{2}-J_{3}+J_{1}).
\end{eqnarray}
Upon entering the region $J_{3}<0$, this phase is further stabilized since the spins coupled via $J_3$ bonds are ferromagnetically aligned in this N\'eel phase.

\subsubsection{Sublattice N\'eel phase}

In the region $J_{2}<0$, when $J_{3}<|J_{2}|$, the free parameter $\alpha$ characterizing phase I is determined to zero implying that all spins within a given unit cell are ferromagnetically aligned. Furthermore, these four-site unit cells form N\'eel order throughout the lattice, and hence, each sublattice is N\'eel ordered [see Fig.~\ref{fig:spinconfigs}(c)], we henceforth refer to this phase as a sublattice N\'eel ordered phase [marked as phase III (orange region) in Fig.~\ref{fig:classical_phase}(a)]. This state is thus characterized by $(k,\alpha)=(\pi,0)$. The ground state energy density can be written as,
\begin{eqnarray}
 E=-\frac{1}{2}(2|J_2|-J_3+J_1).
\end{eqnarray} 
When $J_{3}<0$, the spin confiuration satisfies all the couplings. 

\subsection{Classical Monte Carlo analysis}
\label{sec:cmc}


Since the Luttinger-Tisza approach is not {\it a priori} expected to give the exact ground state spin configuration on a non-Bravais lattice such as ours, and given the fact that our approach is based on a variational ansatz [Eq.~\eqref{en_spin}], we perform classical Monte Carlo simulations to investigate the accuracy of our analysis as well as to study the role of thermal fluctuations. It is worth mentioning that a two-dimensional system of Heisenberg spins with finite-range antiferromagnetic or ferromagnetic interactions cannot feature long-range order at any finite temperature by virtue of the Hohenberg-Mermin-Wagner theorem~\cite{Hohenberg-1967,*Mermin-1966}. Our discussion thus refers to the behavior of order parameters (here spin orientations) at short length-scales, i.e., at distances $r$ less than the correlation radius $\xi$. We consider a system of $1600$ ($=20\times 20\times4$) spins, employ parallel tempering, and carry out simulations at temperatures $T/J_{1}S^{2}=0.001$~\footnote{All the simulations have been performed starting from a high temperature of $T\sim J_{1}$ and reaching down till $T\sim10^{-3}J_{1}$ through slow annealing. We have used $10^{4}$ Monte Carlo steps for thermalization, followed by $10^{5}$ Monte Carlo steps during which measurements are taken every $10$ Monte Carlo steps to ensure uncorrelated results}. We find that in phases II and III both the angles $(\gamma,\alpha)$ [see Fig.~\ref{fig:classical_phase}(b)] lock into the values $(0,\pi)$ and $(0,0)$, respectively, as expected from the Luttinger-Tisza result. In contrast, in phase I, the angle $\gamma$ settles into a value of $\pi$, while thermal order-by-disorder mechanism fails to lift the degeneracy in the angles $\alpha_{1}$, $\alpha_{2}$ and $\alpha_{3}$ [see Fig.~\ref{fig:spinconfigs}(a)] which therefore continue to exhibit a fluctuating behavior with time evolution/Monte Carlo steps~\footnote{We have verified the fluctuating behavior of the angles $\alpha_{1}$, $\alpha_{2}$, and $\alpha_{3}$ down to temperatures $T\sim10^{-5}J_{1}$. At low temperatures, we also perform a restricted metropolis update\textemdash such an update proposes a new spin at random in a conical region about the local field of the old spin~\cite{Zhitomirsky-2008}. Adjusting the size of the conical region gives us control over the acceptance rate of proposed spins which is small at lower temperatures. In our simulation, the acceptance rate was around $50\%$. We have also checked the robustness of our findings by starting from an ordered state obtained by initializing the three angles to $0$ and $\pi$ and observing their evolution with Monte Carlo steps. The results are identical to those obtained by starting from a random configuration, namely that they exhibit fluctuating behavior}. Furthermore, as the temperature $T\to0$ the specific heat $C$ tends to a value less than one, pointing to the important fact that the role of anharmonic fluctuation modes cannot be neglected. Our analysis thus provides evidence for the stabilization of an uncorrelated antiferromagnetic chain phase at \textit{finite temperatures}. This is also reflected in the finding that within the region of parameter space occupied by phase I, the ground state energy obtained from classical Monte Carlo simulations is independent of $J_{2}$ indicating an absence of spin correlations between the different horizontal and vertical chains. Hence, the spin configurations determined from the (numerically) exact classical Monte Carlo simulations are in complete agreement with those determined from the variational ansatz [Eq.~\eqref{en_spin}], validating the classical phase diagram of Fig.~\ref{fig:classical_phase}(a).

Having discussed the classical ground states occupying the $(J_{1}, J_{2}, J_{3})$ parameter space, it is instructive to calculate the magnetic structure factor which  can be experimentally measured in a neutron scattering experiment~\cite{Lovesey-1984} to reveal the signatures of the magnetic ground states. As discussed above, we have found three different magnetic ground states, namely, an uncorrelated antiferromagnetic chain phase, N\'eel phase and sublattice N\'eel phase, depending on the sign and magnitude of the exchange parameters in the
Hamiltonian. Here, we calculate the static (equal-time) spin structure factor
\begin{equation}
S({\bf k})=\frac{1}{N}\sum_{i,j}e^{-\imath{\bf k}\cdot{\bf R}_{ij}}\langle{\bf S}_{i}\cdot{\bf S}_{j}\rangle
\label{stf_0}
\end{equation}
via classical Monte Carlo simulations for three different sets of parameter values. Here, $N$ is the total number of sites, and $i$, $j$ run over all the sites of the lattice, and the nearest-neighbour distance between two sites is set to unity. This implies that the distance between the neighbouring unit cells is three units which makes the periodicity of $S({\mathbf k})$ as $2\pi/3$. In Fig.~\ref{fig_stf} we show the structure factor of the different phases in Fig.~\ref{fig:classical_phase}(a).  We note that in phase I the horizontal and vertical chains are uncorrelated, and thereby have no global ordering in any direction, on average. The perfect antiferromagnetic order in a given chain yields a vanishing contribution when summed over all chains. Hence, we expect a featureless structure factor with the presence of subdued peaks at the zone boundary (corresponding to an ordering vector ${\bf k}$ which is zero upon a statistical averaging) as can be seen in Fig.~\ref{fig_stf}(a). For the  magnetic ground states given by phase II and phase III, we note that a global antiferromagnetic (N\'eel) order should yield peaks at ${\bf k}=(m \frac{\pi}{3}, n \frac{\pi}{3}) $ where $m, n$ are odd integers. However, the peaks at the above mentioned points are modulated due to the form factor of the magnetic ordering in a given unit cell of N\'eel phase (phase II) and subattice N\'eel phase (phase III). These form factors lead to the appearance of blue lines along the diagonal for the phase II making a cross like controur and for phase III a square pattern. The origin of such low intensity contours can be understood by calculating the structure factor for the N\'eel and the sublattice N\'eel phases analytically. An exact expression of the static structure factor for the different classical spin configurations obtained within a Luttinger-Tisza formalism [see Sec.~\ref{grd-st-sec}] can be determined by evaluating Eq.~\eqref{stf_0} analytically. The static structure factor $\mathcal{F}= S^2 |f|^2$ (where $S$ denotes the form factor of the magnetization in a given unit cell and $f$ arises due to global N\'eel ordering) thus obtained is given by
\begin{eqnarray}
&&f=\frac{(1-e^{\imath NLk_x})(1-e^{\imath NLk_y})}{(4N)^2(1+e^{\imath Lk_x})(1+e^{\imath Lk_y})}, \; \; \; L=3a \label{stf} \\
&& S(\mathbf{k})=4(\cos k_x\mp\cos k_y)^2 \label{np}
\end{eqnarray}
for a lattice of $N \times N$ unit cells, and $-(+)$ corresponds to the form factor for the N\'eel (sublattice N\'eel) phase. 

The form factor $S$ is zero along the line $k_x=2 m^{\prime} \pi  \pm k_y$  for phase II and $k_x=(2 m^{\prime} +1) \pi \pm k_y$ for phase III where  $m^{\prime}$ is any integer including zero. Thus, the expected peaks at ${\bf k}=( m \frac{\pi}{3}, n \frac{\pi}{3} )$ will be modulated due to  such lines of zero intensity and it would remove some peaks in structure factor which are expected due to global N\'eel ordering. Figure~\ref{fig_stf}(b) represents the structure factor for the N\'eel order (phase II). We see that the peaks appear at ${\bf k}=( m \frac{\pi}{3}, n \frac{\pi}{3} )$ with $m \ne n$ due to the form factor modulation. On the other hand, such a destructive modulation is absent for the sublattice N\'eel phase and the peaks appear at the expected locations ${\bf k}= ( m \frac{\pi}{3}, n \frac{\pi}{3} )$ with $m$ and $n$ being odd integers [see Fig.~\ref{fig_stf}(c)]. In practice, a particular material may not have the exact symmetry of the lattice we have considered. For example, the unit cell may not be a square as taken here and also the distances between different sites of different neighbors connected by exchange couplings would also be different, in which case the experimentally obtained structure factor would be modified compared to that shown in Fig. \ref{fig_stf}. However, the structure factor in that case can be easily compared by evaluating Eq.~\eqref{stf} with modified lattice parameters, mainly the different values of ${\bf R}_{ij}$.\\

\section{Spin Wave Analysis}
\label{spin-wave}
\subsection{Antiferromagnetic chain phase}\label{sw1}

\begin{figure*}[t]
  \includegraphics[width=1.0\textwidth]{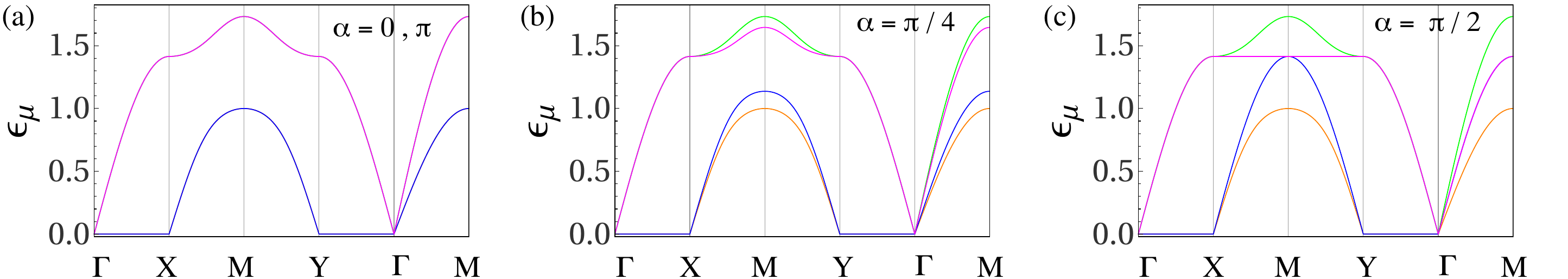}
  \caption{Free magnon spectrum corresponding to Eqs. (21) and
(22) in phase I for the parameter values $(J_2 /J_1 , J_3 /J_1 ) = (1, 2)$ plotted along the high-symmetry
path $\Gamma(0, 0)$, $X (\pi , 0)$, $M(\pi ,\pi)$, and $Y (0, \pi)$ for three choices of degeneracy
parameters, (a) $\alpha=0\; or \;\pi$, (b) $\alpha=\pi/4$ and (c) $\alpha=\pi/2$.}
\label{dis_phase1_h0}
\end{figure*}

The $T=0$ classical antiferromagnetic chain phase [phase I in Fig.~\ref{fig:classical_phase}(a)] has an infinitely degenerate ground state manifold. In particular, on a lattice consisting of $N$ horizontal and $M$ vertical lines [see Fig.~\ref{fig:classical_phase}(b)], there are $N \times M$ independent $\alpha$ parameters each of which can take on values ranging from 0 to $2\pi$. It is thus natural to ask the question whether quantum fluctuations can lift this degeneracy via an order-by-disorder mechanism~\cite{Villain-1980,Shender-1982} and select a unique configuration parameterized by a certain value of $\alpha$. To investigate the role of quantum fluctuations, we carry out a linear spin wave analysis. To this effect, we rotate our coordinate system in such a way that the $z$-axis of the local coordinate system coincides with the axis of the local spin orientation,
\begin{equation}
\hat{S}^{\alpha}_{i}=R_{x}\left( \frac{\pi}{2} \right)R_{z}(\phi_i)\hat{S}^{\alpha^\prime}_{i},~~ \alpha=x,y,z.
\label{rot}
\end{equation}
The Holstein-Primakoff transformation~\cite{Holstein-1940} can now be written as
\begin{eqnarray}
\hat{S}^{z^\prime}_{i,m}& \approx&s-{\hat{a}^\dagger}_{i,m}  \hat{a}_{i,m} \nonumber \\
\hat{S}^{x^\prime}_{i,m}& \approx&\sqrt{\frac{s}{2}}\left(\hat{a}^{\dagger}_{i,m}+\hat{a}_{i,m}\right) \nonumber \\
\hat{S}^{y^\prime}_{i,m}& \approx&\imath\sqrt{\frac{s}{2}}\left(\hat{a}^{\dagger}_{i,m}-\hat{a}_{i,m}\right).
\label{HP1}
\end{eqnarray}
In the above $m=1,2,3,4$ denote the sublattice indices. Within a quadratic approximation to the boson operators we find that a uniform choice of $\alpha$ (when all $N\times M$ values of $\alpha$ are the same) is energetically favorable compared to disordered configurations of $\alpha$ (when all $N\times M$ values of $\alpha$ are different), indicating an order-by-disorder mechanism at work. This lifting of the degeneracy is only {\it partial} as $\alpha$ (which is now the same for all $N\times M$ sites) can still take on any value between $0$ and $2\pi$ yielding the same ground state energy, and the hence there still remain an infinite number of degenerate ground states. However, the magnon spectrum for each value of $\a$ need not be the same and thus we investigate the $\alpha$ dependence of the spin-wave spectrum. In Appendix~\ref{app} we provide the expressions of the resulting Hamiltonian after implementing the Holstein-Primakoff transformation corresponding to Eqs.~\eqref{rot} and~\eqref{HP1}. As expected, the Hamiltonian is invariant under $\mathcal{PT}$-symmetry which is defined as $\mathcal{PT}=\sigma_x \otimes \sigma_0 \mathcal{K}$ (where $\mathcal{K}$ is the complex conjugation operator). In phase I, the magnetic and crystallographic unit cells are identical, as a result of which we get a $8\times 8$ Hamiltonian matrix [see Eq.~\eqref{app_a}] in {\bf k}-space. As a consequence of $\mathcal{PT}$ symmetry we obtain four doubly-degenerate magnon branches shown in Fig.~\ref{dis_phase1_h0} with the following dispersion relations
\begin{eqnarray}
\epsilon(k)_{1,2}&&=\frac{1}{2}\sqrt{p_k \pm \frac{1}{2}\sqrt{Q(0) f_k+ g_k(0)+ h_k  } } \label{e12}\\
\epsilon(k)_{3,4}&&=\frac{1}{2}\sqrt{p_k \pm\frac{1}{2} \sqrt{Q(\a) f_k+ g_k(\a)+ h_k  } } \label{e34}
 \end{eqnarray}
 where $Q(\a)=\frac{2(J_2)^{2}\cos^2\a-(J_{3})^2}{4}$ and
 \begin{eqnarray}
p_k&=&\frac{J_3}{4}(2-\cos k_x -\cos k_y) \\
f_k&=& (\cos(k_x+k_y)+\cos(k_x-k_y)) \\
g_k(\a)&=& (J_{2})^{2} \cos^2 \a(1-\cos k_x -\cos k_y) \\
h_k&=& (J_{3})^{2}(2+\cos(2k_x)+\cos(2k_y) )/8
 \end{eqnarray}
From Eq.~\eqref{e12} we see that there are two modes which are independent of $\a$, and Eq.~\eqref{e34} shows that the other two modes are $\alpha$ dependent. For any given value of $\a$ there are two Goldstone modes originating from the spontaneously broken U(1) symmetry, and we observe the presence of zero-energy modes along the segments $\overline{\Gamma X}$ and $\overline{\Gamma Y}$. The presence of zero-energy modes can be understood from Eqs.~\eqref{e12} and~\eqref{e34} which upon substitution of $k_x (k_y) = 0$ and $k_y (k_x) =k$ yields $\epsilon(k)=0$ and $\epsilon(k)=\sqrt{J_3} \sin k/2$. Hence, there are two doubly degenerate modes along the $k_x=0$ and $k_y=0$ axes out of which one doubly degenerate mode has zero excitation energy and another linearly dispersing in $k$ for small values.

It is worth noting that along the $k_{x}=0$ and $k_{y}=0$ axes we have linear band crossings along the line segments $\overline{\Gamma X}$ and $\overline{\Gamma Y}$ in the Brillouin zone, thus forming Dirac nodal lines. We now discuss how the spin wave spectrum depends on $\alpha$. (i) For $\alpha=0$ or $\pi$: in this case the system has two sublattice N\'eel order. The Hamiltonian is block diagonal [see Appendix~\ref{app}], and each block is $\mathcal{PT}$ invariant which gives rise to two 4-fold degenerate bands [see Fig.~\ref{dis_phase1_h0}(a)]. (ii) For $\alpha=\pi/4$: The block diagonal structure disappears and as a result the two 4-fold degenerate bands split into four 2-fold degenerate bands [see Fig.~\ref{dis_phase1_h0}(b)]. (iii) For $\a=\pi/2$: In this case we have a band touching of the $\a$ dependent bands at the $M$ point which is a consequence of the underlying mirror reflection symmetry about the $k_{x}$, $k_{y}$ and $k_{x}=k_{y}$ axes. We note that there is a Dirac nodal line along the segment $\overline{\Gamma M}$ [see Fig.~\ref{dis_phase1_h0}(c)]. Low energy expansion of dispersion along $\overline{\Gamma M}$ gives
\begin{eqnarray}
\epsilon(\kappa)_{1,2}&&=2(J_3\pm J_2\cos{\alpha})|\sin{\kappa/2}| ~~\\
\epsilon(\kappa)_{3,4}&&=2(J_3\pm J_2)|\sin{\kappa/2}|
\end{eqnarray}
In the above expressions, we have substituted $k_x=k_y=\kappa$. It is evident that the number of independent modes along the $\overline{\Gamma M}$ depends on $\alpha$. For $\alpha= 0, \pi$ there are two doubly-degenerate optical modes as represented by violet and blue lines in Fig.~\ref{dis_phase1_h0}(a). When $\alpha= \pi/4$, the two-fold degeneracy of both the optical modes gets lifted [see Fig.~\ref{dis_phase1_h0}(b)], however, when $\alpha= \pi/2$ only the degeneracy of the lower optical mode gets lifted. The above observation may have experimental relevance in deciding the exchange parameter set for the model Hamiltonian or the selected angle $\alpha$. \\
The appearance of doubly degenerate zero energy modes along the $\overline{\Gamma Y}$ and $\overline{\Gamma X}$ is  reminsicent of the fact that  irrespective of the one-dimensional order along a certain direction, the perpendicular direction can adjust itself free of energy cost. We have also found that the inclusion of anisotropy or magnetic field perpendicular to the spin alignment plane leads to a lifting of the degeneracy of the optical modes along $\overline{YM}$ or $\overline{\Gamma M}$, however, the zero energy mode along the $k_x(k_y)=0$ survives on $\overline{\Gamma Y}$ and $\overline{\Gamma X}$ segments. The inclusion of higher order terms may lift this degeneracy~\cite{Henley-2009} but we expect the zero energy modes at high symmetry points to survive the inclusion of magnon-magnon interaction terms~\cite{Tatsuo1976,Majumdar-2009}.

\begin{figure*}
  \includegraphics[width=1.0\textwidth]{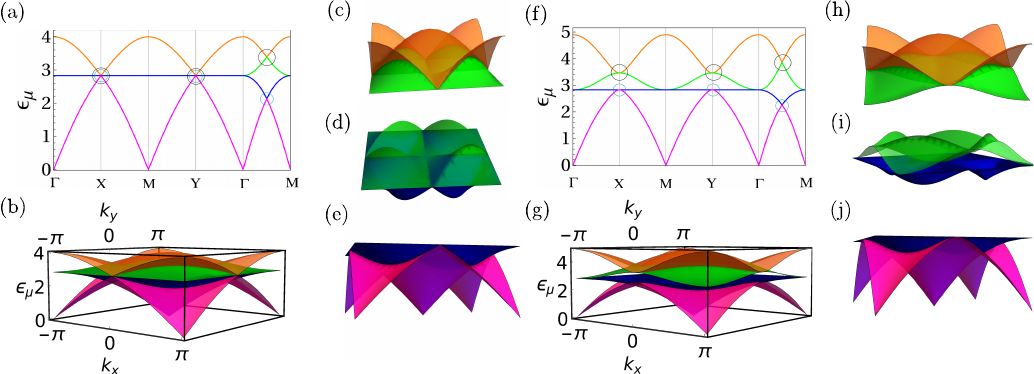}
  \caption{Free magnon spectrum for phase II (N\'eel phase): (a), (b) $(J_2,J_3)=(2,1)$ and (f),(g) $(J_2,J_3)=(3,2)$, (c), (d), (e) and (h), (i), (j) represent the band touchings between different bands, which form the Dirac nodal loops for the two above-mentioned choices of parameters, respectively. Notice that the nodal loop due to touching of the bands denoted by green and blue colors appears for some choices of parameters and is gapped out for other choices. Hence, this nodal loop is not protected by any symmetry whereas the nodal loops due to touching of the upper and lower two bands survive for all choices of parameters, thereby rendering these loops symmetry protected.}
\label{dis_phase2a}
\end{figure*}

\begin{figure}[b]
  \includegraphics[width=1.0\textwidth]{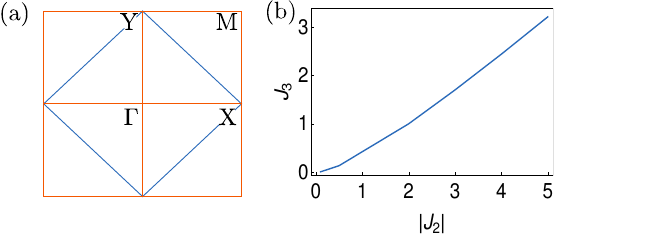}
  \caption{(a) Projection of Dirac nodal loop on the first Brillouin zone. The blue line denotes the symmetry protected nodal loops where as the orange one appers for some specific choices of parameters. (b) denotes the choices of parameters where the additional nodal loop appears which is not protected by any symmetry.}
\label{dis_phase2b}
\end{figure}

\begin{figure}[b]
  \includegraphics[width=1.0\textwidth]{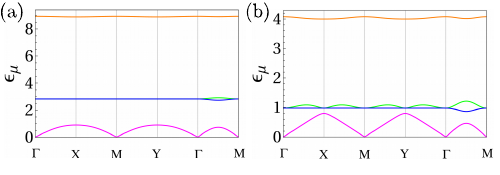}
  \caption{Free magnon spectrum for phase III (sublattice N\'eel phase) for (a) $(J_2,J_3)=(-2,1)$ and (b) $(J_2,J_3)=(-0.8,0.6)$. Unlike N\'eel phase there is no robust Dirac nodal loop in this phase. However similar to N\'eel phase in some choices of parameters a nodal loop (not protected by any symmetry) appears along the zone boundary and $k_x=0$ and $k_y=0$ axes as can be seen from figure (a).}
\label{dis_phase2c}
\end{figure}
\subsection{Phase II and Phase III}
This phase has perfect antiferromagnetic structure which is the same as for the unfrustrated case ($J_3=0$ and $J_1,J_2>0$) studied in Ref.~\cite{Owerre-2018} where the presence of two Dirac nodal loops was found. The magnetic unit cell (8-sites) is twice the size of the crystallographic unit cell (4 sites). In the bipartite representation, the Hamiltonian can be written as in Eq.~\eqref{app_ph2}, and is seen to be block diagonal with each block being $\mathcal{PT}$ invariant. This results in four four-fold degenerate bands shown in Fig.~\ref{dis_phase2a}. Similar to what was found for the unfrustrated model ($J_{3}=0$), we find two nodal loops [marked by blue line in Fig.~\ref{dis_phase2b}(a)], namely, in Figs.~\ref{dis_phase2a}(a) and~\ref{dis_phase2a}(f) we see that the black circled points form one nodal loop while the cyan circled points form the second nodal loop. In the absence of a $J_{3}$ coupling it was shown in Ref.~\cite{Owerre-2018} that the Dirac nodal loops are topologically protected, and that there is a triple band touching at the $\Gamma$ and $M$ points, while in the presence of an additional $J_{3}$ coupling we find a quadruple band touching at the $X$ and $Y$ points as shown in Fig.~\ref{dis_phase2a}(a). Furthermore, we find that a $J_{3}$ coupling leads to the appearance of an additional Dirac nodal loop along the Brillouin zone boundary and also along the $k_{x} (k_{y})=0$ axes [see Fig.~\ref{dis_phase2a}(d) and the orange segment in Fig.~\ref{dis_phase2b}(a)]. However, this additional loop appears only for particular choices of parameters shown in Fig.~\ref{dis_phase2b}(b), and is not protected by any symmetry except at the time-reversal invariant momentum points, i.e., the $\Gamma$ and $M$ points [see Fig.~\ref{dis_phase2a}(f) and Fig.~\ref{dis_phase2a}(i)], where there remains a two-fold degeneracy since the Hamiltonian given by Eq.~\eqref{app_ph2} is invariant under $\mathcal{T}=\imath\sigma_y\mathcal{K}$ operator.

In phase III, whose magnetic structure has an eight site unit cell, we similarly find that the Hamiltonian Eq.~\eqref{app_ph3} is block diagonal with each block being $\mathcal{PT}$ invariant, leading to four four-fold degenerate bands as in phase II. However, the symmetry protected Dirac nodal loops of phase II are not found in phase III, on the other hand, the additional nodal loop along the zone boundary and $k_{x}(k_{y})=0$ axes found for phase II also appears in phase III for the same choice of parameters shown in Fig.~\ref{dis_phase2c}. We note that the surface states for a ribbon geometry in phase II are gapless but the edge states for phase III are gapped.

\begin{figure*}
  \includegraphics[width=1.0\textwidth]{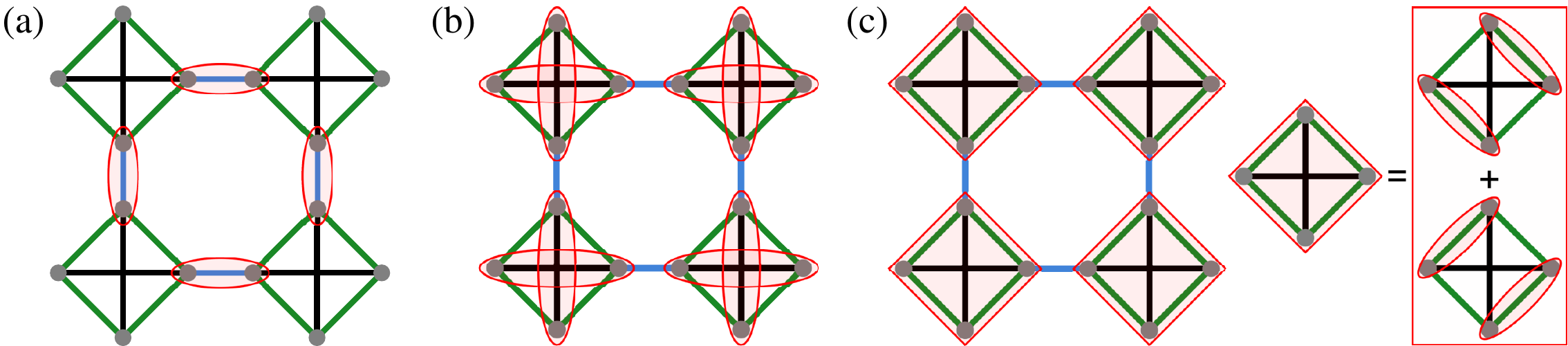}
  \caption{A schematic illustration of the pattern of singlet dimer formations for the parameter regimes of Eq.~\eqref{heisen}, (a) $|J_{1}|\gg(|J_{2}|,|J_{3}|)$ (VBS$_{1}$), (b) $|J_{3}|\gg(|J_{1}|,|J_{2}|)$ (VBS$_{2}$), (c) $|J_{2}|\gg(|J_{1}|,|J_{3}|)$ (Plaquette RVB) wherein the quantum ground state is given by a superposition of two states with dimer formation on the opposite sides of the square as shown in the right panel.}
\label{vbs_fig}
\end{figure*}
\section{Thermal and Quantum Order-by-disorder effects}
\label{fluctuation}
For a two-dimensional system of Heisenberg spins, the Hohenberg-Mermin-Wagner theorem dictates that even at infinitesimally small temperature, the deviation of the spin orientation, i.e., the spin fluctuation, is infinitely large. Thus, the assumption of small fluctuations about the classical ground state, i.e., the harmonic approximation is, in principle, not applicable. Nonetheless, the entropic selection of the dominant magnetic fluctuation tendencies at low temperatures carried out within a harmonic analysis may provide information on the nature of the short range correlations. It is worth mentioning that the harmonic order results can be significantly altered if the anharmonic modes play a decisive role. 

We now proceed to carry out such an analysis for phase I, which at zero temperature is infinitely degenerate being characterized any value of $\alpha\in[0,2\pi]$. However, at finite temperature, the fluctuation of the spins explicitly contribute to the $\alpha$ dependent free energy. This could possibly lead to a lifting of the degeneracy in the parameter $\alpha$ via a thermal order-by-disorder mechanism~\cite{Balla-2019,Henley-1989,Kawamura-1984}. To this effect, we introduce a spin deviation $\theta_{\mathbf R,\mu}\rightarrow{\theta^0}_{\mathbf R,\mu}+\delta\theta_{\mathbf R,\mu}$, where $\theta_{\mathbf R,\mu}$ is the ground state spin configuration of the $\mu^{th}$ sublattice of a unit cell with radius vector $\mathbf{R}$ and $\delta\theta_{\mathbf R,\mu}$ is the deviation from ground state spin configuration. Substituting this in Eq.~\eqref{heisen} and expanding around the ground state up to quadratic order in $\delta\theta_{\mathbf R,\mu}$, in Fourier space we obtain $\hat{\mathcal{H}}=E_{\rm GS}+\hat{\mathcal{H}}_{\rm fluctuation}$ with
\begin{equation}
\hat{\mathcal{H}}_{\rm fluctuation}=\sum_{\mathbf q}\psi^{\dagger}_{\mathbf q} A_{\mathbf q}(\a)\psi_{\mathbf q}
\label{hfluc}
\end{equation}
where $\psi_{\mathbf q}=[\delta\theta_{{\mathbf q},1} \; \delta\theta_{\mathbf q,2} \; \delta\theta_{\mathbf q,3} \; \delta\theta_{\mathbf q,4}]^T$, and $E_{\rm GS}$ is the ground state energy.

In this (harmonic) approximation, the fluctuations can be integrated out in the partition sum, and give rise to a linear-$T$ dependence in the free energy $\mathcal{F}(\alpha,T)$. Following Ref.~\cite{Kawamura-1984}, $\mathcal{F}(\alpha,T)$ can be written as
\begin{equation}
\mathcal{F}(\a,T)= E_{\rm GS}-NT\ln T+T\sum_{\mathbf q\in\rm BZ}\ln(\det A_{\mathbf q}(\alpha)),
\label{eq:free_en}
\end{equation}
where the last term is the $\alpha$ dependent part of the low-temperature entropy density. The state which minimizes this term corresponds to the minimum of the free energy—this is the entropic order-by-disorder selection mechanism discussed in Refs.~\cite{Villain-1980,Shender-1982,Kawamura-1984,Henley-1989}. In the region of the phase diagram occupied by phase I, we find that thermal fluctuations select a value of $\alpha$ equal to $0$ or $\pi$. This is in contrast to classical Monte Carlo result which finds that the order-by-disorder mechanism fails to lift the degeneracy in the angle $\alpha$. Our results thus point to the non-negligible impact of anharmonic order fluctuations which seem to alter the harmonic order picture sharply in favor of an uncorrelated antiferromagnetic chain phase.

Furthermore, the energy of the spin-wave modes is $ \hbar S\epsilon_{\mathbf{q},\mu}(\alpha)$ in the semi-classical description for spins of length $S\gg 1$. Quantum fluctuations then choose the state with the lowest zero-point energy 

\begin{equation}
 \mathcal{E}_{\text{ZP}}(\alpha) =  \sum_{\mathbf{q} \in \text{BZ}} \frac{\hbar S}{2}\epsilon_{\mathbf{q,\mu}}(\alpha).
\label{eq:quantum}
\end{equation}  

The $ \mathcal{E}_{\text{ZP}}(\mathbf{\alpha})$ behaves qualitatively like the the last term of Eq.~\eqref{eq:free_en} and selects the same ordering vectors. In the region of the phase diagram occupied by phase I, we find that at zero-temperature quantum fluctuations select a value of $\alpha$ equal to $0$ or $\pi$. It will be interesting to investigate the impact of anharmonic order terms, which we leave for a future study.

\section{Bond Operator Analysis: Valence bond solid phases}
\label{vbs}

In the extreme quantum limit of $S=1/2$, there arises the possibility of zero-point quantum fluctuations destroying long-range antiferromagnetic orders when the amplitude of the fluctuations becomes of the order of the spin length. Furthermore, the presence of frustrated interactions enhances quantum fluctuations thereby aiding the stabilization of quantum paramagnetic phases such as quantum spin liquids and valence bond solids (VBS). Here, we investigate the effect of quantum fluctuations beyond the spin-wave approximation by employing resonating valence bond variational wave functions. We first investigate our $J_{1}$-$J_{2}$-$J_{3}$ model in parameter regimes where one of the couplings is overwhelmingly stronger compared to the remaining two. In this limiting regime, we note that at zeroth order the strongest bonds will form a singlet or triplet dimer and the locally excited states correspond to singlet to triplet excitations or {\it vice versa} for antiferromagnetic or ferromagnetic bonds, respectively [see Fig.~\ref{vbs_fig}]. The effect of non-zero values of the remaining two (subdominant) couplings is to dynamically create local excitations on the strong singlet and triplet dimer bonds. Hence, an effective Hamiltonian of interacting singlet or triplet bonds between neighboring dimers can be constructed within this approach. This procedure is known to be effective in explaining the low energy physics of interacting spin systems~\cite{Manuel-1998,Doretto-2014,Sachdev-1990,Singh-1999,Kotov-1999,Zhitomirsky-1996,Kumar-2010,Ghosh-2016,Ghosh-2018}. We now discuss the different VBS phases which are found to be realized as variational quantum ground states of the $J_{1}$-$J_{2}$-$J_{3}$ model.  

\subsection{VBS$_{1}$ and VBS$_{2}$ phases}
\label{vbs12}
In the limit when $|J_1|\gg(|J_{2}|,|J_{3}|)$ we have a VBS configuration consisting of dimers on $J_{1}$ type bonds [see Fig.~\ref{vbs_fig}(a)]. The local Hilbert space is four dimensional spanning the singlet ground state and the three triplets as the excited states. Following Ref.~\cite{Sachdev-1990}, we define $\hat{\psi}^{\dagger}_i$ and $\hat{\chi}^{\dagger}_i$ as the creation operators of the singlet and triplet states, respectively, on the $i^{\text th}$ bond within a given unit cell with the accompanying constraint on the dimensionality of the Hilbert space
\begin{eqnarray}
\label{const}
\hat{\psi}^{\dagger}_{i}\hat{\psi}_{i}+\hat{\chi}^{\dagger}_{i,\a}\hat{\chi}_{i,\a}=2S,
\end{eqnarray}
where $i=1$ or 2 corresponds to the two $J_1$ type bonds in a unit cell, and $\alpha=1,2,3$ denotes the three types of triplets in a given dimer. The interaction term between the two neighboring unit cells is obtained by writing down the spin components in terms of the above mentioned valence bond operators. This is achieved by calculating $\langle m |\hat{S}^{\nu}_{\mu} | n\rangle$ where $\mu=1,2$ denotes the two spins in a given dimer, $\nu=x,y,z$ labels the three spin components, and $|m\rangle$ ($|n\rangle$) represents the singlet or triplet states. In general, a spin operator at a given site has the form $\hat{S}_{i}\approx \frac{1}{2}(\hat{\psi}^{\dagger}_{i}\hat{\chi}_{i}+\text{h.c.})$ where $i$ labels a given dimer [see Appendix~\ref{app_vb} for details]. As a result of this transformation we land up with a Hamiltonian which is quartic in the field operators $\hat{\chi}$ and $\hat{\psi}$. We carry out a mean-field decoupling such that the resulting quadratic Hamiltonian [Eq.~\eqref{vb_ham}] separates into singlet and triplet sectors with no mixing terms. The ground state energy is obtained by extremizing the Hamiltonian with respect to the mean-field parameter $N_{i}$ (where $\sqrt{N_{i}}=\langle\hat{\psi}^{\dagger}_{i}\rangle=\langle\hat{\psi}_{i}\rangle$) and the Lagrange multiplier $\lambda$ needed to implement the constraint of Eq.~\eqref{const}, which yields two sets of self consistent equations which are solved. The resulting expression for the ground state energy per unit cell for VBS state in the parameter regime $J_{1}\gg (J_2, J_3)$ is,
\begin{eqnarray}
\nonumber \mathcal{E}_{a}&=(N_{a,1}+N_{a,2})E_{a}-\lambda[(N_{a,1}+N_{a,2})-1]\\&+\frac{3}{2N}\sum_{\bf k}\sum_{i=1}^{2}({\Theta}_{{\bf k},i}-\mathcal{A}_{{\bf k},i}),
\label{vbs1_egs}
\end{eqnarray}
where $a$ labels the VBS$_{1}$ state, $\Theta_{{\bf k},i}$ is the eigenenergy obtained in the triplon sector of the interacting  Hamiltonian Eq.~\eqref{vb_ham} signifying the excitation of the triplet state~\cite{Kumar-2010}, and $\mathcal{A}_{{\bf k},i}$ is the diagonal element of Eq.~\eqref{vb_ham}. A similar procedure can be followed for the VBS$_{2}$ state in the parameter regime $J_3\gg (J_1, J_2)$. For the VBS$_{2}$ state the unit cell has been conveniently chosen as an elementary square. In a similar manner as above, we obtain the ground state energy per square, 
\begin{eqnarray}
\nonumber \mathcal{E}_{b}&=(N_{b,1}+N_{b,2})E_{b}-\lambda[(N_{b,1}+N_{b,2})-1]\\&+\frac{3}{2N}\sum_{\bf k}\sum_{i=1}^{2}({\Pi}_{{\bf k},i}- \mathcal{B}_{{\bf k},i}).
\label{v2_egs}
\end{eqnarray} 
where $b$ labels the VBS$_{2}$ state, and the remaining terms all have identical meaning to that in Eq.~\eqref{vbs1_egs} [see Appendix~\ref{app_vb} for further details]. 

\begin{figure}
\includegraphics[width=1.0\textwidth]{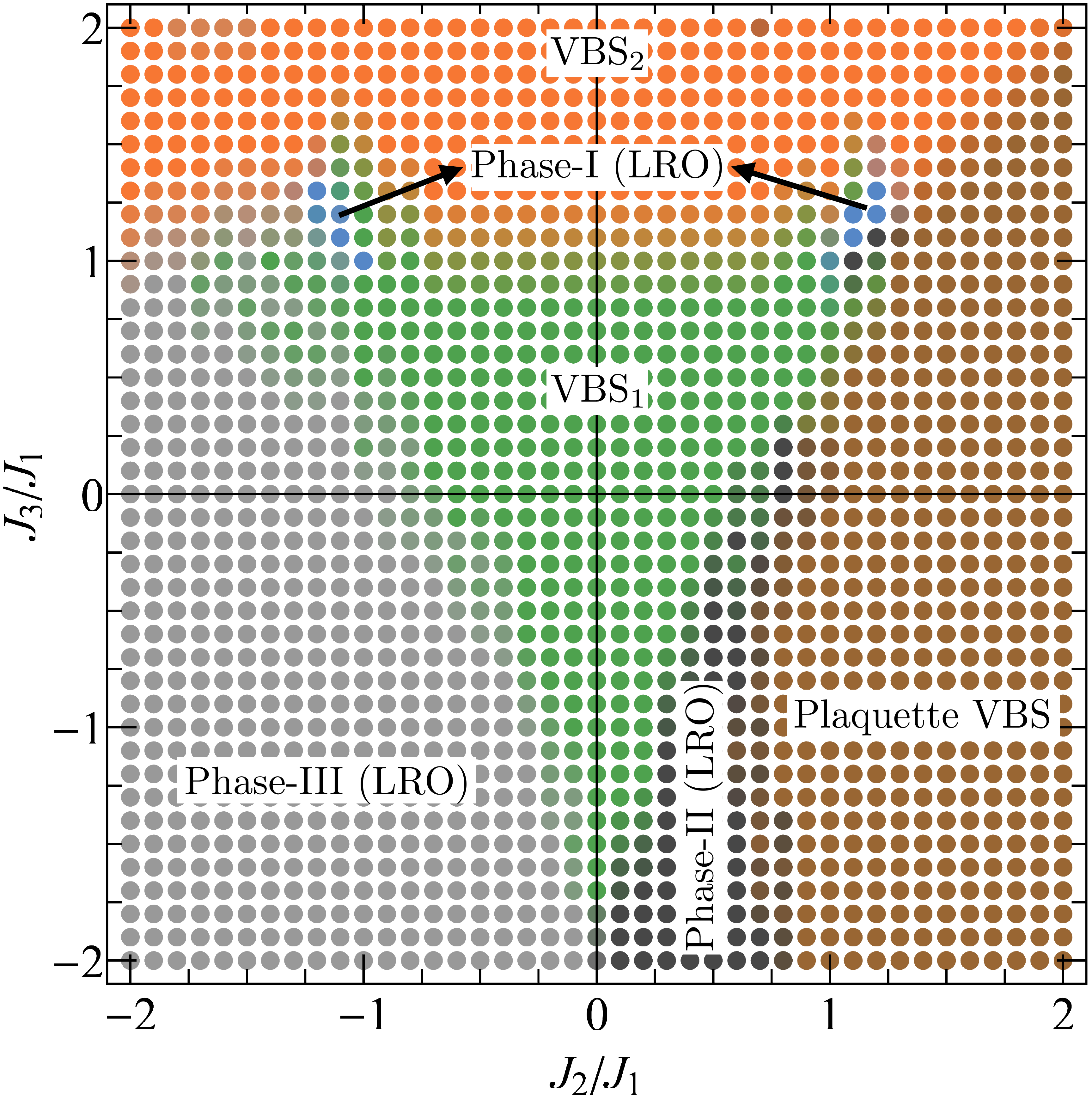}
\caption{Quantum phase diagram of the $S=1/2$ $J_{1}$-$J_{2}$-$J_{3}$ Heisenberg model on the Fisher lattice (Fig.~\ref{fig:lattice}(a)) obtained by a bond operator analysis. We see the appearance of three different VBS phases in addition to the three long-range ordered (LRO) phases also present in the classical phase diagram (Fig.~\ref{fig:classical_phase}(a)).}
\label{vbs_phase_diagram}
\end{figure}

\subsection{Plaquette VBS}
In the parameter regime $(J_2 ,J_3)\gg J_1$ where we expect a plaquette VBS phase [see Fig.~\ref{vbs_fig}(c)], we have two choices of forming two dimers inside a square. More precisely, if we denote the four vertices of a square as V$_{1}$, V$_{2}$, V$_{3}$, and V$_{4}$, then we have two possibilities for dimer formation, namely V$_{1}$\textendash V$_{2}$ and V$_{3}$\textendash V$_{4}$ or V$_{1}$\textendash V$_{4}$ and V$_{2}$\textendash V$_{3}$. The corresponding Hamiltonian is
\begin{equation}
\label{plham}
\hat{\mathcal{H}}_{p}=J_2(\hat{\bf S}_1+\hat{\bf S}_3)\cdot(\hat{\bf S}_2+\hat{\bf S}_4)+J_3(\hat{\bf S}_1\cdot\hat{\bf S}_3+\hat{\bf S}_2\cdot\hat{\bf S}_4)
\end{equation}
The diagonalization of the above Hamiltonian gives the following two lowest energy plaquette singlet states,
\begin{equation}
\label{plwf}
|\Psi_{p,(\pm)} \rangle= \frac{1}{\sqrt{2}} (|\psi_{1,2} \rangle|\psi_{3,4} \rangle\pm |\psi_{1,4}\rangle| \psi_{2,3}\rangle)
\end{equation}
where $|\psi_{i,j}\rangle$ denotes a singlet state formed between the sites $i$ and $j$, $|\Psi_{p,+}\rangle$ is the ground state wavefunction with energy $E^{s}_{+}=-2J_2+\frac{J_{3}}{2}$ and $|\Psi_{p,-}\rangle$ is the first (singlet) excited state with energy $E_{s,-}=-\frac{3}{2}J_3$. Above these states lie the nine triplet states of the plaquette VBS with energies, 
\begin{eqnarray}
E^{t}_{\mu,\mu}=(-J_2+\frac{J_{3}}{2})\delta_{\mu,3}-\frac{J_{3}}{2}(\delta_{\mu,1}+\delta_{\mu,2})
\end{eqnarray}
where $\mu,\nu=1,2,3$. The five quintet states have a degenerate energy $E_d=J_2+\frac{J_{3}}{2}$. To capture the low-energy dynamics we have restricted our analysis to within the singlet-triplet manifold. The low-energy dynamics now includes, in addition to the triplet excited states considered for the VBS$_{1}$ and VBS$_{2}$ states, the singlet excited states. Within this approximation the effective low-energy Hamiltonian for a single plaquette can be written as
\begin{equation}
\hat{\mathcal{H}}_{p}=\sum_{i=\pm}E_{s,i} \hat{\psi}^{\dagger}_{p,i} \hat{\psi}_{p,i} + \sum_{\mu,\nu=1}^{3} E_{t(\mu,\nu)} \hat{\chi}^{\dagger}_{(\mu,\nu)} \hat{\chi}_{(\mu,\nu)}.~~~
\label{plaq_hamiltonian}
\end{equation}
The Eqs.~\eqref{plham}, \eqref{plwf}, and \eqref{plaq_hamiltonian} together with the constraint of Eq.~\eqref{const} provide a complete description of the low-energy spectrum at zeroth order, i.e., in the absence of a $J_{1}$ interaction, leading to isolated plaquettes. The presence of a finite $J_1$ introduces interactions between neighboring plaquettes which induce transitions between different states of Eq.~\eqref{plaq_hamiltonian}. The inter-plaquette interactions are obtained by writing the spin components in terms of the above-mentioned plaquette operators as was done for the VBS$_{1}$ and VBS$_{2}$ states [see Appendix~\ref{app_plaq} for details]. The effective low-energy Hamiltonian for the interacting plaquette-VBS state thus obtained is given in Eq.~\eqref{p_ham} [see Appendix~\ref{app_plaq} for details]. The ground state energy per plaquette is
\begin{equation}
\mathcal{E}_{c}=N_{c,+}(E_{c}-\lambda)+\lambda+ \frac{3}{2N}\sum_{\bf k}\sum_{i=1}^{3}(\Omega_{{\bf k},i}-\mathcal{C}_{{\bf k},i})
\label{plaq_egs}
\end{equation} 
where $\sqrt{N_{c,+}}=\langle\hat{\psi}^{\dagger}_{p,+}\rangle=\langle\hat{\psi}_{p,+}\rangle$, $\Omega_{{\bf k},i}$ is the eigenenergy obtained in the triplon sector of the interacting plaquette Hamiltonian [Eq.~\eqref{p_ham}].

\begin{figure}
\includegraphics[width=0.8\textwidth]{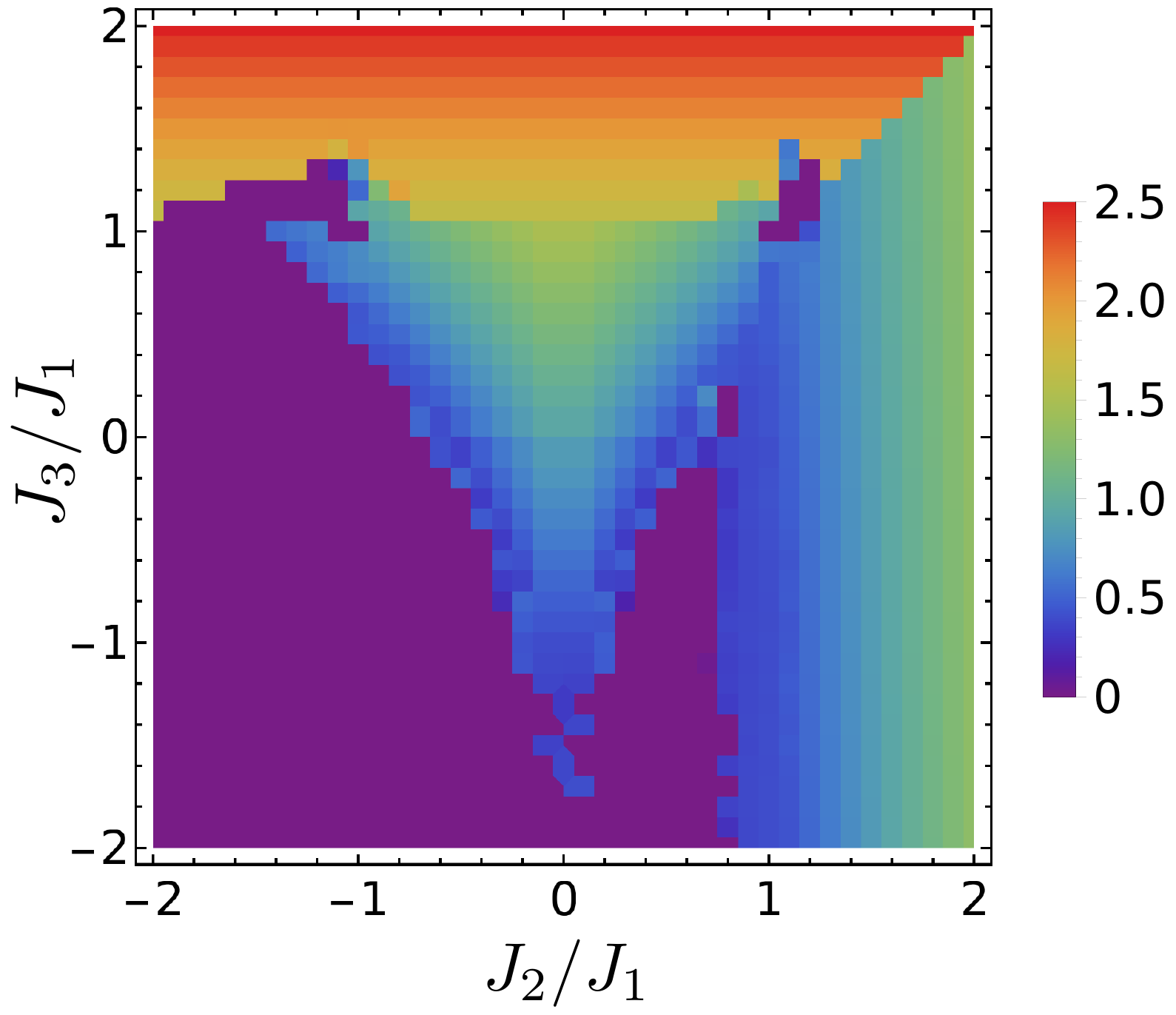}
\caption{Density plot of triplet excitation gap obtained by plaquette and bond operator analysis.}
\label{triplon_gap}
\end{figure}

\begin{figure}
\includegraphics[width=0.85\textwidth]{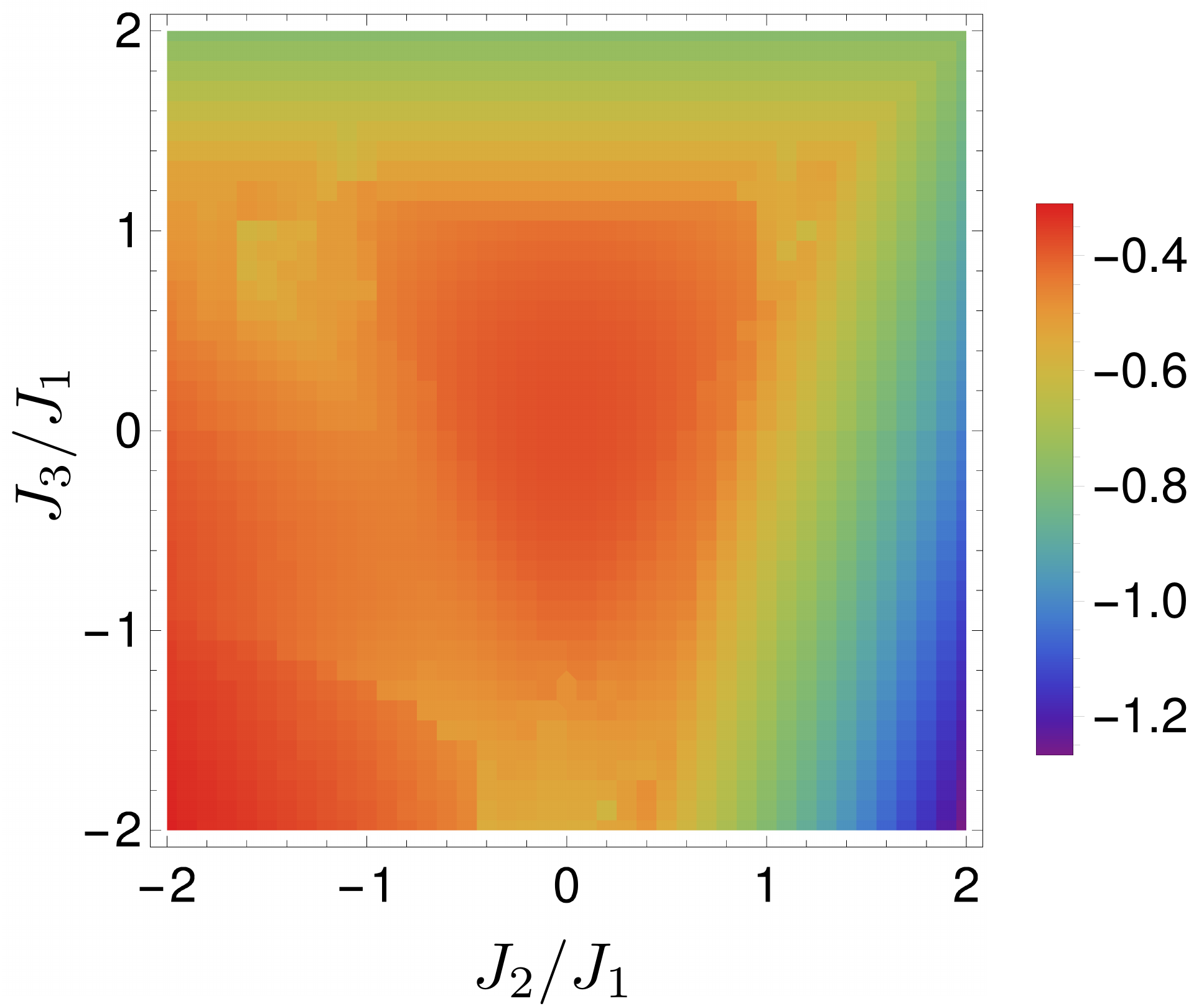}
\caption{Density plot of ground state energy obtained by plaquette and bond operator analysis.}
\label{gr_ene}
\end{figure}

Employing the expressions [Eq.~\eqref{vbs1_egs}, Eq.~\eqref{v2_egs} and Eq.~\eqref{plaq_egs}] of the ground state energies of the three phases we map out the resulting phase diagram. The most salient feature of our phase diagram is the appearance of three quantum paramagnetic phases, namely, a plaquette VBS, and two other types of dimer ordered states dubbed VBS$_{1}$ and VBS$_{2}$ as shown in Fig. \ref{vbs_phase_diagram}. The classical region of existence of the uncorrelated  AF chain phase is now found to be divided into two region under the influence quantum fluctuations. For $J_1 \gg (J_2,J_3)$, the VBS$_1$ state is stabilized while for $J_3 \gg (J_1,J_2)$, the VBS$_2$ state is stabilized. At the phase boundary of these two VBS states, a long range ordered (LRO) state is found to be stabilized in a sliver of parameters space as inferred by the vanishing singlet-triplet excitation gap. We label this long range ordered state as phase I (LRO) and it is characterized by the ordering vector ${\mathbf Q}=(0,k_y),(k_x,0)$. The ordering wave vectors of the three LRO phases shown in Fig. \ref{vbs_phase_diagram} are determined by the momenta associated with the singlet-triplet gaps of the VBS phases that vanish at the N\'eel-VBS quantum phase transition

Another interesting aspect of the quantum phase diagram is that the N\'eel phase (labeled as phase II in Fig. \ref{fig:classical_phase}(a)) gives was to a plaquette-RVB state irrespective of the sign of the $J_{3}$ coupling. On the other hand, the sublattice N\'eel phase (labeled as phase III in Fig. \ref{fig:classical_phase}(a)) is largely immune to quantum fluctuations, and is referred to as phase III (LRO) in Fig. \ref{vbs_phase_diagram}. Apart from these phases, we find another long range ordered phase in a small region for negative $J_3$ sandwiched between VBS$_{1}$ and plaquette VBS phases and label it as phase II (LRO) in Fig. \ref{vbs_phase_diagram}. For the quantum paramagnetic phases such as VBS$_1$, VBS$_2$ and plaquette RVB, the singlet-triplet excitation gap is used as a measure of determining their stability as ground states with respect to inclusion of higher order corrections. For the three long range ordered phases shown in Fig. \ref{vbs_phase_diagram}, the respective wave vectors which characterize these phases correspond to the vanishing of the singlet-triplet excitation gap of the VBS phases. 

\begin{figure}
\includegraphics[width=1.0\textwidth]{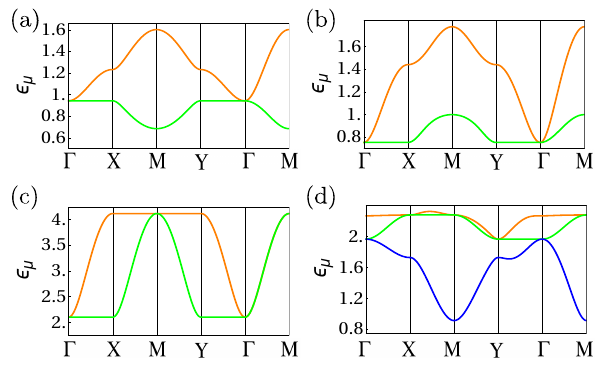}
\caption{Dispersion spectra for different VBS phases for representative values of $(J_2/J_1,J_3/J_1)$ (a) VBS$_1 (\pm 0.5,0.3)$, (b) VBS$_1 (\pm 0.5,0.7)$, (c) VBS$_2 (\pm 1.0,1.5)$ and (d) Plaquette VBS $ (1.5,1.1)$.}
\label{vb_dispersion}
\end{figure}

The singlet-triplet excitation gap for the various VBS phases is shown in Fig. \ref{triplon_gap}. In Fig. \ref{gr_ene}, we prsesent a density plot of the ground state energy density for all the phases in the $J_2-J_3$ plane. It indicates that the ground state energy of the LRO phases is higher compared to the VBS phases which highlights the role of quantum fluctuations in stabilizing quantum paramagnet phases. In Fig. \ref{vb_dispersion}, we present the excitation spectrum for different VBS phases at a few representative parameter values. In Fig. \ref{vb_dispersion}(a) and Fig. \ref{vb_dispersion}(b) we notice that depending on the value of $J_3$, the minimum of the dispersion occurs at either at the $M$ or the $\Gamma$ point. The local minima at the $M$ point changes to a local maxima upon increasing the value of $J_3$. This points to the fact that when $J_3  < J_2$ it is easier to create excitations through $J_2$ bonds which require an antiferromagnetic ordering between the dimers formed on $J_1$ bonds. This virtual antiferromagnetic ordering shows up as minima at the $M$ points [see Fig.~\ref{vb_dispersion}(a)]. However, when  $J_3 > J_2$ such virtual antiferromagnetic ordering is not favorable as indicated by the local maxima at the $M$ point [see Fig.~\ref{vb_dispersion}(b)]. In Fig. \ref{vb_dispersion}(c) we present the excitation spectrum for VBS$_2$ which shows a maxima at the $M$ point. The excitation spectrum for plaquette RVB is shown in Fig.~\ref{vb_dispersion}(d) is a gapped quadratic dispersion with minima at the $M$ points.  

The above results obtained within bond operator formalism suggests that the large degeneracy in the disordered antiferromagnetic chain phase in the classical limit may not lead to a ground state degeneracy in the exact quantum limit as indicated by the stability of the VBS$_2$ phase due to large values of the singlet-triplet excitation gap as shown in Fig. \ref{triplon_gap}. On the other hand, the N\'eel phase which is known to be susceptible to quantum fluctuations is gives way to a plaquette VBS state. It is to be noted that the singlet-triplet excitation gap for VBS$_1$ state is smaller compared to that of the VBS$_2$ state. In fact, Fig. \ref{triplon_gap} suggests that the singlet-triplet excitation gap gradually decreases with decreasing $J_3$. While the uncorrelated antiferromagnetic chain phase and N\'eel phase yield to quantum paramagnetic states under quantum fluctuations, remarkably the sublattice N\'eel phase is quite stable as evident from Fig. \ref{vbs_phase_diagram}. This  may be attributed to the fact that the number of nearest neighbour bonds with ferromagnetic alignment is three times than the number of bonds with antiferromagnetic alignment. The other possible explanation is that each square plaquette can be thought of as a large spin with magnitude of $4S$ which protects it from  quantum fluctuations~\cite{Ghosh-2019a}. Finally, we note that similar observations of a plaquette VBS, and competing magnetic phases on a variant of the model, namely, the square kagome lattice Heisenberg model have previously been made~\cite{Lugan-2019,Nakano-2013,Rousochatzakis-2013,Morita-2018,Richter-2009,Ralko-2015}.
\section{Discussion}
\label{discussion}

We have investigated the ground state phase diagram of the Heisenberg model on the Fisher lattice in the presence of first neighbor $J_1$, second neighbor $J_2$, and third neighbor $J_3$ Heisenberg couplings, as a route towrads providing a magnetic model of a two dimensional layer of the Hollandite lattice. At the classical level, a Luttinger-Tisza analysis shows that the phase diagram is host to three different phases, namely, (i) an uncorrelated antiferromagnetic chain phase wherein each horizontal and vertical  chain has perfect one-dimensional antiferromagnetic order but the relative orientations between any two chains is not fixed at $T=0$~\cite{Anderson-1956,McClarty-2015,Balla-2020}. This uncorrelated antiferromagnetic chain phase exists for $J_2 \geqslant J_3$ and antiferromagnetic $J_2,~ J_3$ {\it only}. Furthermore, there exist two different N\'eel phasse depending on the sign of $J_2$. For antiferromagnetic $J_2$, we find a N\'eel phase (phase II) with the four spins within a unit cell being antiferromagnetically ordered. On the other hand, for ferromagnetic $J_{2}$, we find that the four spins within a unit cell are aligned ferromagnetically with such a cluster of ferromagnetic spins forming N\'eel order, namely the sublattice N\'eel order (phase III).

We have investigated the role of thermal and quantum order-by-disorder effects. Interestingly, our classical Monte Carlo analysis finds that the uncorrelated antiferromagnetic chain phase survives at finite temperature, i.e., the order-by-disorder mechanism fails to lift the degeneracy. On the other hand, a harmonic order analysis of quantum fluctuations reveals an order-by-disorder transition by selecting a common angle $\alpha$ between all the one dimensional chains. We find that although the zeroth order energy in the spin wave approximation is identical for each $\alpha$, the details of the spin wave spectrum depend on the value of $\alpha$, e.g., the spectrum could be linear or quadratic for different values of $\alpha$, and the number of zero energy modes depends on $\alpha$. Interestingly, quantum fluctuations (within a harmonic order treatment) lift the degeneracy and select $\alpha=0, \pi$. For specific choices of $\alpha$, the spin-wave spectrum shows Dirac nodal lines along $\overline{\Gamma X}$ and $\overline{\Gamma Y}$ segments. The spin-wave spectrum for phase II reveals the presence of three Dirac nodal loops out of which two are symmetry protected and do not depend on the value of $J_3$.

Finally, we have employed a bond operator formalism to analyze the model Hamiltonian beyond the spin-wave wave approximation. The analysis for spin $S=1/2$ shows that most of the classical phases except the sublattice N\'eel phase give way to different types of valence bond states. The region of parameter space classically occupied by the uncorrelated antiferromagnetic chain phase is stabilized into VBS$_1$ dimer order with an appreciable singlet-triplet excitations gap. This VBS$_1$ phase appears mostly for large positive values of $J_3$.  In a triangular shaped region around the centre in $J_2-J_3$ plane, a VBS$_2$ dimer state is stabilized. The N\'eel phase largely gives way to a plaquette VBS  state, while the sublattice N\'eel phase is found to be stable under quantum fluctuations within the bond operator formalism.

We expect that our study would set the stage for further investigations into the magnetic phases on the Hollandite lattice~\cite{Crespo-2013a, Mandal-2014,Amber-2015,Liu-2014}. The experimental realization of two dimensional layers of $\alpha$-MnO$_2$ (if possible) might serve as a platform to confirm the existence of some of the phases that have been found here. A possible extension of the present study is to include a coupling between such two dimensional layers yielding a three dimensional model of magnetism in Hollandite lattice. As a future endeavor, it would be interesting to study the spin $S=1/2$ quantum phase diagram employing state-of-the-art numerical quantum many-body frameworks such as pseudofermion functional renormalization group~\cite{Reuther-2010,Iqbal-2016} and variational quantum Monte Carlo methods~\cite{Capriotti-2001} which have already been applied on the square~\cite{Hu-2013} and other and other two- and three-
dimensional lattices~\citep{Iqbal-2018,Chillal-2020}. In particular, for $S=1/2$ there exists the likely possibility of quantum spin liquid(s) occupying a finite region of parameter space~\cite{Baskaran-2009}. It will be worthwhile to carry out a projective symmetry group classification~\cite{Wen-2002} of U(1) and $\mathds{Z}_{2}$ quantum spin liquids on the Fisher lattice. The resulting {\it Ans\"atze} and the competition between them could then be studied either by combining the projective symmetry group classification framework with a functional renormalization group approach~\cite{Hering-2019} or employing variational Monte Carlo on the corresponding Gutzwiller projected wave functions supplemented by a few Lanczos steps~\cite{Iqbal-2011b,Iqbal-2013,Iqbal-2014,Iqbal-2015,Iqbal-2016a,Iqbal-2018a}. In particular, it would be important to investigate their stability and energetic competitiveness with the valence-bond solid orders, similar to what has been done on the kagome lattice~\cite{Iqbal-2011a,Iqbal-2012}. Finally, we would like to mention that an interesting avenue for further exploration would be to possibly destabilize ferromagnetic order on the Fisher lattice which could potentially give rise to a plethora of nematic orders as has been found in the square lattice~\cite{Shannon-2006,Iqbal-2016b}

\section*{Acknowledgements}
Y.I. thanks B. Dabholkar and K. Penc for helpful discussions.
Y.I. acknowledges financial support by the Science and Engineering Research Board, Department of Science and Technology, Ministry of Science and Technology, India through the Startup Research Grant No.~SRG/2019/000056 and MATRICS Grant No.~MTR/2019/001042. This research was supported in part by, the National Science Foundation under Grant No.~NSF~PHY-1748958, the Abdus Salam International Centre for Theoretical Physics (ICTP) through the Simons Associateship scheme funded by the Simons Foundation, the International Centre for Theoretical Sciences (ICTS), Bengaluru, India during a visit for participating in the program “Novel phases of quantum matter” (Code: ICTS/topmatter2019/12) and “The 2nd Asia Pacific Workshop on Quantum Magnetism” (Code: ICTS/apfm2018/11). The Monte Carlo simulations were performed at SAMKHYA: High Performance Computing Facility provided by Institute of Physics, Bhubaneswar.

\appendix

\section{Hamiltonian for spin-wave spectrum}
\label{app}
In the  appendix, we explicitly write the Hamiltonian which is used to find the spectrum of spin wave excitations.

\subsection{Phase I}
For phase I, which we refer to as the uncorrelated antiferromagnetic chain phase, the Hamiltonian matrix is,
\begin{eqnarray}
H_{\bf k} &=& 
\begin{bmatrix}
A({\bf k}) & B({\bf k}) \\
B({\bf k}) & A({\bf k})
\end{bmatrix} . 
\label{app_a}
\end{eqnarray}
The basis vector is chosen to be $  \hat{\tilde{\c}}_{\bf k}=[\hat{\c}_{\bf k},\hat{\c}^\dagger_{-{\bf k}}]^T$, with $\hat{\c}_{\bf k}=(\hat{a}_{{\bf k},1}, \; \hat{a}_{{\bf k},2}, \; \hat{a}_{{\bf k},3}, \; \hat{a}_{{\bf k},4})$. In the above, $A_{{\bf k}}$ and $B_{{\bf k}}$ are $4 \times 4 $ matrices 
\begin{eqnarray}
A_{\bf k} &=& 
\begin{bmatrix}
a & b & 0 & -c \\
b & a & -c & 0 \\
0 & -c & a & b \\
-c & 0 & b & a
\end{bmatrix}
\end{eqnarray}
\begin{eqnarray}
B_{\bf k} &=& 
\begin{bmatrix}
0 & c & b_x({\bf k}) & -b \\
c & 0 & -b & b_y({\bf k}) \\
b^*_x({\bf k}) & -b & 0 & c \\
-b & b^*_y({\bf k}) & c & 0
\end{bmatrix}.
\end{eqnarray}
Among the various parameters that appear in the above two equations $a, b, c $ are constants with
 $a=\frac{1+ J_3}{2}, b= \frac{J_2}{4} (1- \cos \a), c= \frac{J_2}{4} (1+\cos \a)$. $b_{(x/y)}({\bf k})$
is given below,
\begin{eqnarray}
&& b_x({\bf k})=- \frac{1}{2}(J_3+e^{\imath k_x}),  b_y({\bf k})=-\frac{1}{2}(J_3+e^{-\imath k_y}).~~~~~
\end{eqnarray}

\subsection{Phase II}

For the N\'eel phase, the  Hamiltonian becomes an $8\times8$  hermitian matrix due to its four sublattice
structure and the presence of global antiferromagnetic order. The basis vector which is used to define the Hamiltonian is  $\hat{\tilde{\c}}_{\bf k}=[\hat{\c}_{\bf k},\hat{\c}^\dagger_{-{\bf k}}]^T$, with $\hat{\c}_{\bf k}=(\hat{a}_{{\bf k},1}, \; \hat{a}_{{\bf k},2}, \; \hat{a}_{{\bf k},3}, \; \hat{a}_{{\bf k},4}, \; \hat{b}^\dagger_{-{\bf k},1}, \; \hat{b}^\dagger_{-{\bf k},2}, \; \hat{b}^\dagger_{-{\bf k},3}, \; \hat{b}^\dagger_{-{\bf k},4})$, where $\hat{a}_{\bf k}$ and $\hat{b}_{\bf k}$ respectively denotes the up spin and down spin in momentum space. The  Hamiltonian is obtained as,
\begin{eqnarray}
H_{\bf k} &=& I_{2 \times 2}\otimes
\begin{bmatrix}
A({\bf k}) & B({\bf k}) \\
B({\bf k}) & A({\bf k})
\end{bmatrix},
\label{app_ph2}
\end{eqnarray}
where $A_{\bf k}$ and $B_{\bf k}$ are given below.
\begin{eqnarray}
A_{\bf k} &=& 
\begin{bmatrix}
d & 0 & -J_3 & 0 \\
0 & d & 0 & -J_3 \\
-J_3 & 0 & d & 0 \\
0 & -J_3 & 0 & d
\end{bmatrix},
\end{eqnarray}
\begin{eqnarray}
B_{\bf k} &=& 
\begin{bmatrix}
0 & J_2 & e^{\imath k_x} & J_2 \\
J_2 & 0 & J_2 & e^{\imath k_x} \\
e^{-\imath k_x} & J_2 & 0 & J_2 \\
J_2 & e^{-\imath k_x} & J_2 & 0
\end{bmatrix}.
\end{eqnarray}
 In the above $d=J_1+2J_2-J_3$ and denotes the ground state energy per plaquette in phase II.
 
 \subsection{Phase III}
For the sublattice N\'eel phase, the basis vector used to define the Hamiltonian matrix for each momentum is identical to phase II. The Hamiltonian contains a few additional parameters. The Hamiltonian has the following expression,
\begin{eqnarray}
H_{\bf k} &=& I_{2 \times 2}\otimes
\begin{bmatrix}
A({\bf k}) & B({\bf k}) \\
B({\bf k}) & A({\bf k})
\end{bmatrix}
\label{app_ph3}
\end{eqnarray}
where $A_{\bf k}$ and $B_{\bf k}$ are given below.
\begin{eqnarray}
A_{\bf k} &=& 
\begin{bmatrix}
d & J_2 & -J_3 & J_2 \\
J_2 & d & J_2 & -J_3 \\
-J_3 & J_2 & d & J_2 \\
J_2 & -J_3 & J_2 & d
\end{bmatrix},
\end{eqnarray}
\begin{eqnarray}
B_{\bf k} &=& 
\begin{bmatrix}
0 & 0 & e^{\imath k_x} & 0 \\
0 & 0 & 0 & e^{\imath k_x} \\
e^{-\imath k_x} & 0 & 0 & 0 \\
0 & e^{-\imath k_x} & 0 & 0
\end{bmatrix} ,
\end{eqnarray}
 where $d=J_1-2J_2-J_3$ denotes the ground state energy per plaquette in phase III.

\section{Valence Bond Operator Analysis}
 \label{app_vb}
Here, we provide the detailed procedure followed in the bond operator formalism. First, we give the  definition of spins in terms of the field operators $\hat{\psi}$ and
$\hat{\chi}$ associated with the singlet and triplet excitations~\cite{Sachdev-1990,Kumar-2010}, respectively.
\begin{eqnarray}
&&\hat{S}_{1,\a}\approx \frac{1}{2}(\hat{\chi}^\dagger_{\a}\hat{\psi}+\text{h.c.}),\quad \hat{S}_{2,\a}\approx -\frac{1}{2}(\hat{\chi}^\dagger_{\a}\hat{\psi}+\text{h.c.})
\label{psiphi}
\end{eqnarray}
In the above, the subscript $1,\;2$ refers to two spins within a dimer and $\alpha=x,y,z$ represents the three components of the spin. In defining the above transformation we have restricted ourselves up to quadratic order in the fields. The above definitions can be used to write down the effective Hamiltonian in terms of the field operators. The effective Hamiltonian contains a Lagrangian multiplier $\lambda$ in order to ensure that the magnitude of total spin of a given dimer is $2S$. It is straightforward to observe that the use of Eq.~\eqref{psiphi} yields quartic terms in field operators. Mean-field type decomposition has been used to reduce these quartic terms into appropriate quadratic terms in singlet and triplet sectors, while neglecting the mixing between them. Furthermore, as we are interested in finding the excitations due to triplets over singlet condensation, we introduce 
$ \sqrt{N_{a,i}}=\left<\hat{\psi}^\dagger_{a,i}\right>=\left<\hat{\psi}_{a,i}\right>$ as the singlet occupation number which is used to define the zeroth order condensate energy $\tilde{E}_{a}$. Here `$a$' denotes the VBS$_1$ configuration and `$i=1,2$' refers to two dimers within an unit cell. Similar definition holds for VBS$_2$ which is labelled by the subscript `$b$'. Hence, we can write the effective Hamiltonian for the VBS$_1$ as
\begin{equation}
 \hat{\mathcal H}_{a} = \tilde{E}_{a}+\frac{1}{2}\sum_{\bf k}{\hat{\phi}^\dagger}_{\bf k,\a}H_{{\bf k},{a}}\hat{\phi}_{\bf k,\a}.
\label{vb_ham}
\end{equation}
In the above expression, the singlet condensate energy $\tilde{E}_{a}$ is
\begin{eqnarray}
\tilde{E}_{g,a}&&=(N_{a,1}+N_{a,2})E^s_{a} -\lambda(N_{a,1}+N_{a,2}-1) \nonumber \\
                 &&~ -\frac{3}{2N}\sum_{\bf k}\sum_{i=1,2}A_{{\bf k},a,i}
\end{eqnarray}
where, $A_{{\bf k},a,i}$  represents the $i^{\text th}$ diagonal element of $H_{{\bf k},a}$ and $E^s_{a}=-3J_1/4$ is  energy of the singlet states per plaquette. The second term in  Eq.~\eqref{vb_ham} refers to triplet excitations. The basis vector  used to obtain Eq.~\eqref{vb_ham} is $\hat{\phi}_{{\bf k},\a}=[{\hat{\xi}}_{\bf k},\hat{\xi}^\dagger_{-\bf k}]^T$, with $\hat{\xi}_{\bf k}=(\hat{\chi}_{a,1,\alpha,\bf k}, \hat{\chi}_{a,2,\alpha,{\bf k}})$, where $\alpha$ denotes different states within triplet sector. It is  clear that $H_{{\bf k},a}$ is a $4 \times 4 $ matrix which can be written as, 
\begin{eqnarray}
\hat{\mathcal H}_{{\bf k},a} &=&
\begin{bmatrix}
V_{{\bf k},a}+ D_{a} & V_{{\bf k},a} \\
V_{{\bf k},a} & V_{{\bf k},a}+D_{a} 
\end{bmatrix}
\label{vbmat},
\end{eqnarray}
where $V$ and $D$ are $2 \times 2$ matrices
\begin{eqnarray}
V_{{\bf k},a} &=&
\begin{bmatrix}
-\frac{J_3}{2}N_{a1}\cos 2k_x & J_2\sin k_x \sin k_y \\
J_2 N_{a12}\sin k_x \sin k_y & -\frac{J_3}{2}N_{a2}\cos 2k_y 
\end{bmatrix},~~~~~
\end{eqnarray}

\begin{eqnarray}
D_{a} &=&
\begin{bmatrix}
E^t_{a,\a}-\lambda & 0  \\
0 & E^t_{a,\a}-\lambda 
\end{bmatrix}.
\end{eqnarray}
In the above we have used $N_{a,12}=\sqrt{N_{a1}N_{a2}}$. $E^t$  refers to energy of triplet states with $E^t_{a,\a}=J_1/4 $ i.e., all the triple states are degenerate in energy.  To obtain the corresponding representations for VBS$_2$ we use the singlet and triplet state energies as $ E^s_{b}=-3J_3/4,~ E^t_{b,\a}=J_3/4 $. All other expressions of $V_{{\bf k},a}$ as given in
Eq.~\eqref{vbmat} will be replaced by $V_{{\bf k},b}$ which is given below,

\begin{eqnarray}
V_{{\bf k},b} &=&
\begin{bmatrix}
-\frac{J_1}{2}N_{b1}\cos k_x & 0 \\
0 & -\frac{J_1}{2}N_{b2}\cos k_y
\end{bmatrix}.
\end{eqnarray}

\section{Plaquette Operator Analysis}
\label{app_plaq}
For the plaquette VBS state  represented in  Fig. \ref{vbs_fig}(c), the spin operators are written as~\cite{Doretto-2014}, 
\begin{eqnarray}
\label{plfer}
\nonumber \hat{S}_{\delta,\a}&& \approx c_{\delta,\mu}(\hat{\chi}^\dagger_{\mu,\a}\hat{\psi}_{+}+\text{h.c.})\\
&&~ + d_{\delta,\nu}(\hat{\chi}^\dagger_{\mu,\a}\hat{\psi}_{-}+\text{h.c.}),
 \end{eqnarray}
 where $\a=x,y,z$, $\mu=1,2,3$ denotes the nine triplets and $\delta=1,2,3,4$ denotes site indices inside a plaquette. A summation over the repeated
indices is implied in Eq.~\eqref{plfer}. The matrix $c_{\delta,\mu}$ and $d_{\delta,\nu}$ are given below.
\begin{eqnarray}
c_{\delta,\mu}=
\frac{1}{\sqrt{6}}\begin{bmatrix}
\frac{1}{\sqrt{2}} & 0 & 1 \\
0 & \frac{1}{\sqrt{2}} & -1 \\
-\frac{1}{\sqrt{2}} & 0 & 1 \\
0 & -\frac{1}{\sqrt{2}} & -1 
\end{bmatrix}, d_{\mu,\nu}=
\frac{1}{2}\begin{bmatrix}
0 & 1 & 0 \\
1 & 0 & 0 \\
0 & -1 & 0 \\
-1 & 0 & 0
\end{bmatrix}.~~~~~
\end{eqnarray}
To derive the effective Hamiltonian as obtained for the valence bond singlet states in Appendix~\ref{app_vb} in Eq.~\eqref{vb_ham}, we follow a procedure similar to that explained before Eq.~\eqref{vb_ham}. After doing elementary algebra we obtain the effective Hamiltonian in this case
\begin{eqnarray}
 \hat{\mathcal H}_c &&= \tilde{E}_c +\frac{1}{2}\sum_{{\bf k}}\hat{\phi}^\dagger_{{\bf k},\a}H_{\bf k}\hat{\phi}_{{\bf k},\a} + \sum_{\bf k}(E^s_{-}-\lambda) \hat{\psi}^{\dagger}_{{\bf k}-} \hat{\psi}_{{\bf k}-}.~~~~~~
\label{p_ham}
\end{eqnarray}
 The first term $\tilde{E}_{c}$ in the above equation corresponds to ground state condensate energy per plaquette. For the second term, we have used the basis vector as, $\hat{\phi}_{{\bf k},\a}=[\hat{\xi}_{\bf k},\hat{\xi}^\dagger_{-{\bf k}}]^T$, with $\hat{\xi}_{\bf k}=(\hat{\chi}_{1,\a,{\bf k}} , \hat{\chi}_{2,\a,{\bf k}} , \hat{\chi}_{3,\a,{\bf k}})$. Below we provide explicit expressions of the various terms present in Eq.~\eqref{p_ham}. First we provide $\tilde{E}_c$,
\begin{eqnarray}
\label{ec}
\tilde{E}_{c} && = N_{c+}E^s_{+}-\lambda(N_{c+}-1) -\frac{3}{2N}\sum_{{\bf k}}\sum_{i=1,2,3}\mathcal{C}_{{\bf k},i}~~~~~
\end{eqnarray}
where we have used $\sqrt{N_{c+}}=\left<\hat{\psi}^\dagger_{+}\right>=\left<\hat{\psi}_{+}\right>$ and $\lambda$ is the Lagrange multiplier to satisfy the constraint of  total angular momentum of the dimer to be $2S$.  $E^s_{\pm}$ is the plaquette singlet state energy for the state $| \Psi_{\pm} \rangle$ with $E^s_{+} = -2J_2 + \frac{J_3}{2}, E^s_{-}=-\frac{3J_3}{2}$.  $\mathcal{C}_{{\bf k},i}$ is the `$i$'th diagonal element of $H_{{\bf k}}$. The $6 \times 6$ Hamiltonian matrix $H_{{\bf k}}$ in the second term of Eq.~\eqref{p_ham} is
\begin{eqnarray}
\label{hkm}
\hat{\mathcal H}_{{\bf k}} &=& 
\begin{bmatrix}
W_{{\bf k}}+D_c & W_{{\bf k}} \\
W_{{\bf k}} & W_{{\bf k}}+D_c 
\end{bmatrix}
\end{eqnarray}
where
\begin{eqnarray}
\label{cksk}
W_{{\bf k}} &=& \frac{N_{c+}}{3}
\begin{bmatrix}
\frac{-1}{2}\cos k_x & 0 & \frac{-\imath}{\sqrt{2}} \sin k_x \\
0 & \frac{-1}{\sqrt{2}}\cos k_y & \frac{-\imath}{\sqrt{2}}\sin k_y \\
\frac{\imath}{\sqrt{2}}\sin k_x  &\frac{\imath}{\sqrt{2}} \sin k_y  & \cos k_x + \cos k_y 
\end{bmatrix}~~~~~
\end{eqnarray}

\begin{eqnarray}
D_c &=&
\begin{bmatrix}
E^t_{1,\a}-\lambda & 0 & 0 \\
0 & E^t_{2,\a}-\lambda & 0 \\
0 & 0 & E^t_{3,\a}-\lambda 
\end{bmatrix}
\end{eqnarray}
where $i, \a =1,2,3$ denotes the nine triplet states. After diagonalization, the fluctuation due to triplon excitation contributes to the ground state energy and the final expression for the ground state energy can be written as given in Eq.~\eqref{plaq_egs}. We note that the above analysis has been carried out for $J_2 > J_3$ where $| \Psi_{+} \rangle$ is the ground state
and $| \Psi_{-} \rangle$ is the first excited states. When $J_3 > J_2$ with $J_1=0$, $| \Psi_{-} \rangle$ becomes the ground state and $| \Psi_{+} \rangle$  becomes the first excited state. Thus, there are parameter regimes where analogous analysis needs to be followed considering $| \Psi_{-} \rangle$ as the ground state.  However after doing that  we find that the final energy  obtained in the former case, i.e., when $| \Psi_{+} \rangle $, is the ground state is lower compared to the case when $| \Psi_{-} \rangle $ is the ground state. To obtain the energy expression when $| \Psi_{-} \rangle $ is the ground state one needs to replace the `$-$' subscript of the third term  in Eq.~\eqref{p_ham} by `$+$', `$+$' subscript in Eq.~\eqref{ec} by `$-$'. The expression of $W_{\bf k}$ as given in  Eq.~\eqref{cksk} have the following form,
\begin{eqnarray}
W_{{\bf k}} &=& 
\frac{-N_{c-}}{4}\begin{bmatrix}
\cos k_y & 0 & 0 \\
0 & \cos k_x & 0 \\
0 & 0 & 0 
\end{bmatrix}
\end{eqnarray}
\\


%

\end{document}